\documentclass[11pt,epsf]{article}
\usepackage[textures]{epsfig}

\newcommand{\sect}[1]{\setcounter{equation}{0}\section{#1}}
\newcommand{\subsect}[1]{\subsection{#1}}

\newcommand{\appen}[1]{\setcounter{equation}{0}\appendix{#1}}

\newfont{\frak}{eufm10 scaled\magstep1}
\newfont{\extra}{msbm10 scaled\magstep1}

\newcommand{\extr}[1]{\mbox{\extra #1}}

\newcommand{\C}{\extr C}
\newcommand{\F}{{\bf F}}
\newcommand{\M}{{\bf M}}
\newcommand{\N}{\extr N}
\newcommand{\R}{\extr R}
\newcommand{\Z}{\extr Z}
\newcommand{\p}{\partial}
\newcommand{\al}{\alpha}
\newcommand{\bb}{\beta}
\newcommand{\la}{\lambda}

\newcommand{\ro}{\rho}

\def\p{\partial}

\def\k{\kappa}
\def\skappa{\sqrt{\kappa}}
\def\mskappa{\sqrt{-\kappa}}

\def\be{\begin{equation}}
\def\ee{\end{equation}}
\def\bea{\begin{eqnarray}}
\def\eea{\end{eqnarray}}

\def\>#1{{\bf #1}}                 

\def\group{SO_{\kappa}(3)}

\def\sgroup{\overline{SO}_{\kappa}(3)}
\def\algebra{so_{\kappa}(3)}
\def\salgebra{\overline{so}_{\kappa}(3)}
\def\sphere{S_{\kappa}^{2}}
\def\vers{\,{\rm{vers}}_{\k}}
\def\varaa{\varepsilon_{\k}^{l}}
\def\varaaa{\varepsilon_{\k}^{l-1}}
\def\varaao{\varepsilon_{0}^{l}}
\def\vara{{\varepsilon}_{\k}}
\def\ketm{\,|\vara,\bb, m\rangle}
\def\ketma{\,|\varaa,\bb, m\rangle}

\def\ketmp1{\,| \vara,\bb, m+1\rangle}
\def\ketmm1{\,|\vara,\bb, m-1\rangle}
\def\ketw{\,|\vara,\bb, -l\rangle}
\def\ketW{\,|\vara,\bb, l\rangle}
\begin{document}

\begin{center} 
{\LARGE{\bf{Landau quantum systems: \\[1ex]
an approach based on symmetry}}}
\end{center}

\bigskip\bigskip

\begin{center} 
J. Negro, M.A. del Olmo and A. Rodr\'{\i}guez-Marco 
\end{center}

\begin{center} 
{\sl Departamento de F\'{\i}sica Te\'orica, Universidad de Valladolid, \\
E-47011, Valladolid, Spain.}\\ {e-mail: jnegro@fta.uva.es, olmo@fta.uva.es} 
\end{center}

\vskip 1.5cm
\vskip 1.5cm

\bigskip

\begin{abstract}
We show that the Landau quantum systems (or integer quantum Hall 
effect  systems) in a plane, sphere or a hyperboloid,  can be
explained in a  complete meaningful way from group-theoretical
considerations concerning the  symmetry group of the corresponding
configuration space. The crucial point in our development is the
role played by locality and its appropriate mathematical framework,
the fiber bundles. In this way the Landau levels can be understood as the
local equivalence classes of the symmetry group. We develop a unified
treatment that supplies the correct geometric way to recover the
planar case as a limit of  the spherical or the  hyperbolic quantum systems
when the curvature goes to zero. This is an interesting case where a
contraction procedure gives rise to nontrivial cohomology starting from a
trivial one. We show how to reduce the quantum hyperbolic Landau
problem to a Morse system  using horocyclic coordinates. An
algebraic analysis of the eigenvalue equation allow us to build ladder operators
which can help in solving the spectrum under different boundary conditions.
\end{abstract}
\vskip 3.5cm



\vfill   \eject
\sect{Introduction\label{introduccion}}

The planar Landau levels  arise in the frame of
quantum mechanics (QM) when a charged particle evolves under the 
influence of an external constant magnetic field perpendicular to the
plane  \cite{landau}. Landau quantum systems can also be generalized to  
other surfaces with  a normal stationary magnetic field.  In this way, the
spherical and hyperbolic Landau systems have been also studied
\cite{dunne}, but there is still a lack of a comprehensive  characterization
of these systems from the point of view of their symmetry. We will try to
fill this gap here by a systematic study of such a kind of quantum
systems based on their spatial symmetries.   

A wide  theoretical as well as experimental  activity   has been deployed
around two-dimensional (2D) quantum systems of charged particles in the 
last two decades. In particular, the quantum Hall effect \cite{klitzing}, 
2D systems of electrons subjected to  strong external
magnetic fields at very low temperatures, has received a lot of attention
for its interesting and surprising properties \cite{prange90}. The first step
in the understanding of such effects is simply to undertake the study of
quantum Landau  systems. All that  encourages us to revise the Landau
problem from the optics of symmetry. 

The relevant symmetry group of the magnetic field in the planar
Landau system is the  Euclidean group $E(2)$.  In the same way, the
associated symmetry groups of the spherical and hyperbolic systems are
$SO(3)$ and $SO(2,1)$, respectively.  Moreover, the configuration spaces 
of such Landau systems  (i.e., sphere, plane and hyperboloid) can be
seen as homogeneous spaces  of their corresponding symmetry groups.
Thus, we will set up the following project: i) to carry out a simultaneous
study of these classes  of  Landau systems by using a unifying formalism
that will allow us to compare directly the features of all of them; ii) to
characterize clearly the elements of the Landau systems that can be
explained exclusively in terms of group theoretical arguments.  We shall
develop  the first point of this program thoroughly starting from the definition of a general
symmetry group up to the final solutions of the wave equations. In particular,
we will understand the correct way in which the planar Landau
quantum system can be seen as a limit of the spherical or hyperbolic systems
when the surface curvature vanishes. This question deserves a careful attention
because it displays how  trivial extensions can originate a nontrivial one,
much in the same way as the Poincar\'e group leads to the extended Galilei
group (which is essential to describe the mass of  nonrelativistic systems).

With respect to the second point, up to now the Landau systems were defined by means of
Schr\"odinger equations, and their symmetries played a complementary role as a help to
solve the spectrum. Now, in our viewpoint the key object is the
symmetry  group itself from which to develop a certain canonical procedure to get  the
quantum Landau systems. We will see that the main clue to deal with this problem is the 
concept of locality. Thus, as Bargmann and Wigner already stressed
\cite{BW48}, local representations (or locally operating representations) 
of Lie groups of space-time transformations  \cite{H76,COS85} 
constitute a relevant ingredient in QM. Here, we shall show that local
representations of the symmetry groups are the right approach to describe
the Landau systems (as it was done with the Euclidean group \cite{COS85,olmo01} 
or with the Maxwell groups in \cite{NO90}) providing us at
the same time with the minimal coupling rule of interaction with the
external magnetic field. In conclusion, we can state that, from the symmetry
point of view, local equivalence is the responsible for the classification of
different Landau levels defined on any surface. To show the way this is
realized, and its physical implications, will be one of the main objectives of
the present work. 

The natural framework to write down  local
representations is the language of fiber bundles, so we shall briefly
consider this point along our exposition, but leaving the technical details
to the quoted references in order to shorten the length of this paper. 

The organization of the work is as follows.  Section \ref{so3kappa} is 
devoted to introduce a general group (in fact a one-parameter family of
groups) that includes the three symmetry groups mentioned above, together
with its homogeneous spaces. It is also considered the central extensions
of such groups 
that we will call `magnetic' groups. In Section 
\ref{localrealigroup} we characterize the local representations  of this
general group that will be relevant to define in Section \ref{schrodingerequations} the
Schr\"odinger wave equations for  quantum systems supporting this
symmetry group.  Some basic facts related with the formulation of gauge
invariant potentials under  local realizations in the framework of fiber
bundles are presented in Section
\ref{gaugepotential}. They will allow us to give a group-theoretical
justification of the minimal electromagnetic coupling. In Section
\ref{uirso3kappa} we classify the elementary systems associated to the
magnetic groups in the sense of Wigner \cite{wigner39}, i.e., an elementary
quantum system is associated to a  unitary irreducible realization of the
symmetry group (here we will restrict ourselves to bounded
representations). Afterwards, we decompose the local representations of
the magnetic groups in terms of their elementary  systems  in order to  get
the energy  spectrum and eigenfunctions of the corresponding Landau
quantum systems.  In Section \ref{horocycliccoordinates} we present the
variable separation of the hyperbolic Landau system using the horocyclic
coordinates of $SO(2,1)$.  In this way we reduce the quantum Landau problem to 
a system of a particle  moving in a Morse potential allowing to understand the
continuous spectrum of the hyperbolic system (this question was previously
addressed but only at a classical level). In the following Section we
construct ladder operators connecting eigenstates of consecutive eigenvalues
of the spectrum (for $\k\neq 0$ such operators have not been considered previously up to
our knowledge.)  These ladder operators have some interesting properties:
i) they satisfy essentialy cubic commutation relations; ii) connect
the Landau systems to isotropic oscillators on constant curvature
surfaces; and iii) allow to derive directly the spectrum even when the
wavefunctions obey different boundary conditions (this is the case of the `moving states'). Finally 
Section
\ref{conclusiones} displays the main results  in a more physical language
together with some general remarks  and comments. Some appendices have
been added in order to have a work as selfcontained as possible: Appendix A
gives a short review of local realizations; Appendix B deals with central
extensions of Lie groups and Lie algebras; Appendix C characterizes the
local representations of the magnetic groups, and Appendix D supplies the
basic  elements of fiber bundles and gauge theories.

\sect{Symmetry groups of Landau quantum systems}\label{so3kappa} 

The first step to achieve our program is to propose a unified notation 
by introducing a Lie group, denoted by $SO_{\kappa}(3)$ \cite{ball93},
involving the three aforementioned symmetry groups, and a homogeneous
space which also includes as particular cases the three types of
two-dimensional surfaces where the quantum Landau systems will live.

\subsect{Symmetry groups of constant magnetic fields}

As we mentioned in the introduction the suitable symmetry groups of our 
Landau systems are $SO(3)$, $E(2)$ and  $SO(2,1)$. They can be dealt with
in a more compact way by defining a one-parameter family of Lie groups
$\group$, with $\k$ a real parameter, whose Lie algebra, $\algebra$, is
generated by the infinitesimal (Hermitian) generators $ J_{01}, J_{02}$ and
$J_{12}$ satisfying the following Lie commutators: 
\begin{equation} \label{kso3}
 [J_{01},J_{02}]= i\, \kappa\,  J_{12},\qquad  
[J_{12},J_{01}]= i\,  J_{02},\qquad  
[J_{12}, J_{02}]= - i\, J_{01}.  
\end{equation} 
When $\k$ is nonzero it can be rescaled to $+1$ or $-1$, whence we have
three representative values: $+1,0,-1$. If $\k=+1$ we recover the Lie 
algebra $so(3)$; for $\k=0$ we have the Lie algebra $e(2)$ of the
two-dimensional Euclidean group $E(2)$;  and finally, when $\k=-1$, we get
$so(2,1)$.
The quadratic Casimir of $\algebra$ is
\begin{equation}\label{casimir}
C_{\kappa}=J^{2}_{01} + J^{2}_{02} + {\kappa} J^{2}_{12} .
\end{equation}

The group $\group$ admits a linear action in the ambient space $\R ^{3}$, 
leaving invariant the quadratic form  
$\langle x,x\rangle_{\k} = 
x^{2}_{0}+\kappa x^{2}_{1} +\kappa x^{2}_{2}, \ x\in{\R}^3$. The matrix
representation (that explains the index notation) of the generators is  
\be\label{matrixrepresentation}
J_{01}= i\, (-\k E_{01} +E_{10}),\quad
J_{02}= i\,(-\k E_{02} +E_{20}), \quad 
J_{12}= i\,(- E_{12} +E_{21}), \label{a}
\ee
where the $3{\times} 3$ matrices $E_{ij}$ are defined by  $(E_{ij})_{kl} =
\delta_{ik}\delta_{jl}$, $i,j,k,l=0,1,2$.
In this representation, the orbit of the point $x_0=(1,0,0)$ is the 2D
surface $S^2_{\kappa}$ of equation \be\label{hipersuper}
x^{2}_{0}+\kappa x^{2}_{1} +\kappa x^{2}_{2}=1.
\ee
This surface is diffeomorphic to the homogeneous space
$\group /SO(2)$, where  $SO(2)$ is the isotropy group of $x_0$ 
spanned by $J_{12}$ (the only compact generator of $\algebra$ for any
value of $\k$). For  $\k=+1,0,-1$, the surface 
(\ref{hipersuper}) is the 2--sphere, $S^2$,  the Euclidean plane, ${\mit\Pi}^2$,
and the  hyperboloid, $H^2$,  respectively. So, the parameter $\k$
appearing in the commutation rules (\ref{kso3}) can also be interpreted as the
curvature of $S^2_\k$. In particular, if $\k =0$ the metric  $\langle
x,x\rangle_{\k}$ is degenerate  and the homogeneous space is flat (for more
details see \cite{olmo2}).

The In\"on\"u--Wigner contraction process \cite{inonu} that allows to get
$E(2)$ from 
$SO(3)$ or $SO(2,1)$, is equivalent in our framework to take simply 
$\k= 0$  in (\ref{kso3})  \cite{aops97}. 
This replacement can be interpreted geometrically as  a
deformation where the curvature radius $R=1/\skappa$
($R=1/\sqrt{-\k}$ for the hyperboloid) goes to $\infty$. In this way the
Euclidean plane  becomes the limit of the sphere or the hyperboloid in
equation (\ref{hipersuper}).

A useful chart of $\sphere$ is  given by polar geodesic coordinates
\cite{herranz}. Let us consider again the point $x_{0}=(1,0,0)$ of
$\sphere$, then any other point $x$ of $\sphere$ is parameterized by the pair
$(r,\theta )$  according to the following action of $\group$
\begin{equation}
x=e^{-i\,\theta\, J_{12}}e^{-i\, r\, J_{01}}x_0.
\end{equation}
If $\k $ is positive, $(r,\theta)\in (0,\pi/\sqrt{\kappa})\times  (0,2\pi)$, 
while for $\k$ zero or negative $(r,\theta)\in (0,\infty)\times (0,2\pi)$.
So,  this  chart covers $\sphere$ except the two ``poles" (taking the
point  $x_0$ as the ``north pole" and placing the ``south pole" at the infinity
for the non-compact cases, ${\mit\Pi}^2$ and $H^2$) and the meridian joining
them.  The explicit expression of this coordinate system is
\begin{equation}\label{polares}
x^0 =  \cos \skappa r,  \quad
x^1 = \sin \skappa r  \cos\theta /\skappa, \quad
x^2 = \sin \skappa r\sin\theta /\skappa \  . 
\end{equation}
With this convention, the contracted 2D plane $\sphere$ in the limit $\k
\to 0$ is given by $x^0 =1$, that is, we have chosen the contraction around
the north pole $x_{0}=(1,0,0)$.

The fundamental vector fields associated to the basis generators of  
$\algebra$ that correspond to the action of $\group$ on 
$\sphere$ are
\bea\label{fields0}
&&J_{01}(r,\theta)=-i\, \cos\theta\, \p_r +i\, \skappa\,
\frac{\sin\theta}{\tan\skappa r}\, \p_{\theta},\nonumber\\ 
&&J_{02}(r,\theta)=-i\, \sin\theta\, \p_r -i\, \skappa\,
\frac{\cos\theta}{\tan\skappa r}\, \p_{\theta},\label{campos-0} \\  
&&J_{12}(r,\theta)=-i\, \p_{\theta}.\nonumber
\eea 
These formulae are valid for any value of $\k$. Note that for $\k < 0$ we 
have hyperbolic functions since 
$$
\cos\skappa r=\cosh\mskappa r, \qquad
\frac{\sin\skappa r }{\skappa} = \frac{\sinh\mskappa r}{\mskappa},
$$
while for $\k=0$
$$
\lim_{ k\to 0}\cos\skappa  r=1,\qquad 
\lim_{ k\to 0}\frac{\sin\skappa r }{\skappa} =r .
$$
Hence, when $\k = 1 ,-1$ or $0$  expressions (\ref{campos-0}) give the 
usual vector fields  of $so(3)$, $so(2,1)$ or $e(2)$, respectively. In
particular, for  $\k = 0$ we  immediately obtain the Euclidean fields on
the plane: 
$$
J_{01}(r,\theta) {=} {-} i\cos \theta\, \p_r  
{+} i \frac{\sin\theta}r\, \p_{\theta},\quad
J_{02}(r,\theta) {=} {-}i\sin\theta\, \p_r  
{-} i\frac{\cos\theta}r\, \p_{\theta},\quad
J_{12}(r,\theta) {=} {-} i \p_{\theta}.
$$
Notice that in the Euclidean  limit  ${J}_{01}$ and ${J}_{02}$ become the
generators of translations along the cartesian axes $X$ and $Y$
respectively, while ${J}_{12}$ corresponds to the generator of rotations
with respect to the
$Z$--axis; in this case they are usually  denoted by $P_1,\ P_2$ and $J$.

The invariant measure in $S^2_{\k}$ is given, up to a constant 
factor, by
\be\label{measure}
\sigma =  \frac{\sin \skappa r}{\skappa}\, dr\wedge d\theta .
\ee
In the limit $\k \to 0$ we recover the usual
Euclidean measure 
$\sigma =  r\,  dr\wedge d\theta \ .$

There are other  (group) coordinates (for instance, parallel geodesic or
horocyclic) that have interest to analyze  particular aspects. However,
polar geodesic coordinates \cite{herranz} are more suitable to handle  bases
of eigenfunctions of $J_{12}$ for which the realization (\ref{fields0}) 
is well adapted.

\subsect{Magnetic groups  of Landau systems}\label{gruposmagneticos}

If a physical system has a symmetry group $G$, in QM its symmetry
transformations are described by  projective representations in
the space of rays, or by  representations up to a factor in the
associated  Hilbert space \cite{wigner39, bargmann54}. Such
representations can be obtained by means of  true representations of 
an extended group $\overline G$ that Wigner called ``quantum mechanical
symmetry group'' (Appendix A). 

In our case (see Appendix B)  $\overline G$ is a central extension of
(the universal covering of) ${SO}_\k(3)$ by $\R$ which will be denoted
$\overline{SO}_\k(3)$ and in the following it will be referred to as the
family of ``magnetic groups". The basis 
$\{ \overline{J}_{01}, \overline{J}_{02}, \overline{J}_{12},B\}$ 
of $\overline{so}_\k(3)$, the Lie algebra of
$\overline{SO}_\k(3)$, includes a new generator $B$ corresponding  to the
central extension. The commutators of $\overline{so}_\k(3)$ are given
by    
\begin{equation} \label{exso3b} 
{
[\overline{J}_{01},\overline{J}_{02}]=i\, \kappa\, \overline{J}_{12} + i\, {B},
\quad  
[\overline{J}_{12},\overline{J}_{01}]=i \,\overline{J}_{02},\quad 
[\overline{J}_{12},\overline{J}_{02}]=-i \,\overline{J}_{01} , \quad  [.,B]=0 } . 
\end{equation}
From (\ref{exso3b}) it is easy to see at the level of Lie algebras that 
only when $\k=0$ the extension is nontrivial, giving in this case the
extended Euclidean algebra  $\overline {e}(2)$ \cite{COS85}. 

The group law of $\overline{ SO}_\k(3)$ can be obtained from the Lie
algebra (\ref{exso3b}), but we shall never need it; for us it will be 
enough to work with the infinitesimal generators having in mind its 
physical meaning.  The second order Casimir  is \begin{equation}
\label{casex} {\overline C}_{\kappa}=\overline{J}^{2}_{01} +
\overline{J}^{2}_{02} +  {\kappa}\,
\overline{J}^{2}_{12}+2{B}\overline{J}_{12} . \end{equation}

The homogeneous space 
$\sphere$ can also be expressed as  $ \sphere\approx \group
/{SO}(2)=\sgroup/ (\overline{SO}(2)\otimes \R)$, where
$\overline{SO}(2)$ is a two-fold covering of ${SO}(2)$, and $\R$ is
the group generated by $B$. Since the  extension is central,  the action of the subgroup
$\langle B\rangle$ on $\sphere$ is trivial.

\sect{Local representations of magnetic groups\label{localrealigroup}}

In  Appendix A the reader can find a brief review about the theory of local
representations and in Appendix C  how to build up
the local representations of the magnetic groups $\overline{SO}_\k(3)$,  which
 are the suitable ones to describe the quantum symmetries of ${SO}_\k(3)$.
We shall present in the following the results necessary for our development.

The local  representations (\ref{locreal}) of the basis
generators of $\overline{ so}_\k(3)$ are given by 
Hermitian differential operators that have the general form
\begin{equation}
\label{locgen}
 \overline{X}_j(x) =  X_j(x) +  W_j(x), \qquad
 {B} = - \, \beta ,
\end{equation}
where $\overline{X}_j\in 
\{ \overline{J}_{01}, \overline{J}_{02}, \overline{J}_{12}\}$; 
$X_j(x)$ are the fundamental fields (\ref{fields0}), $W_j(x)$ are real
functions, and $\beta$ is a real number that represents the central generator
$B$ and specifies the factor system of the realization. The final explicit
expressions (obtained along the lines of Appendix C) for the infinitesimal
generators (\ref{locgen}), using polar coordinates (\ref{polares}), are: 
\bea\label{campos-Oexb} 
&&{\overline J}_{01}= \,J_{01}(r,\theta) -  \bb\vers r\,  
\frac{\sqrt{\k}\sin\theta}{\sin\sqrt{\k}\, r},\nonumber\\[1ex]  
&&{\overline J}_{02}=\, J_{02}(r,\theta) +\bb\vers r\,
\frac{\sqrt{\k}\cos\theta} {\sin\sqrt{\k}\, r}, \\[1ex]  
&&{\overline J}_{12}=\, J_{12}(r,\theta), \nonumber\\[1ex] 
&&{ B}=-\bb,\nonumber 
\eea
where  the fields $J_{..}(r,\theta)$ are given in (\ref{fields0}).  
We have also introduced a general versine function \cite{Abr72}
$\vers r = \frac{1}{\k}(1-\cos\skappa r)$ that
has a well defined limit
\be
\lim_{\k \to 0}\vers r = r^{2}/2. 
\label{vs}
\ee

We shall remark some important features of the above realization 
(\ref{campos-Oexb}).  i) First of all, it is instructive to check that
expressions (\ref{campos-Oexb}) indeed satisfy the commutation rules
(\ref{exso3b}). ii) The (extended) fields (\ref{campos-Oexb}) are
smooth around the north pole $x_0$, so that they act on functions also
differentiable there.  iii) The  main point to stress here is that, as it is
detailed in Appendix C, each class of local equivalence for the extended
fields of the form (\ref{locgen}) satisfying (\ref{exso3b}) is characterized by
$\bb$, where  $\bb \in  \R$ if $\k=0$, or $2\bb/\k \in  \Z$ if $\k\neq 0$.
The  reason underlying the discretization of $\bb$ is the same as with
respect to the spin: only half integer values are allowed in the (projective)
representations of $SO(3)$ (or the discrete series of $SO(2,1)$). Other values
of
$\bb$ would lead us to a representation of the algebra, not of the group.

The fields  (\ref{campos-Oexb}), defined up to a
local equivalence, determine a trivial extension for $\k \neq 0$.  
When $\k \to 0$  their expressions reduce to
\bea 
&&{\overline J}_{01} = -i\, \cos \theta \p_r  
+ i\,\frac{\sin\theta}r\, \p_{\theta}  
-  \bb\, \frac{r\sin \theta}2 ,\nonumber\\[1ex]
&&{\overline J}_{02} = -i\, \sin\theta \p_r  
- i\,\frac{\cos\theta}r\, \p_{\theta} 
+ \bb\, \frac{r\cos \theta}2 ,\label{exeu}\\[1ex] 
&&{\overline J}_{12} = -i\,\p_{\theta},\nonumber\\[1ex]  
&&{ B}=-\bb,\nonumber  
\eea
but now the extension becomes nontrivial.
Following the arguments of Appendix C, the limit $\k \to 0$ from
(\ref{campos-Oexb}) to (\ref{exeu}) must be done  having in mind that
$2\bb/\k \in\Z$. If we keep 
$\bb =\bb_0$ fixed, this contraction is discrete since $\k=2\bb_0/n$, $n\in \N$, 
and $n \to \infty$.

\sect{Schr\"odinger equations for  Landau systems}\label{schrodingerequations}

Once obtained the local realizations of $\group$, we can characterize the
quantum  elementary  systems behaving under this type of symmetry
transformations. Thus, we will assume that the support space of the local
realization contains  the Hilbert space of wavefunctions of the system. By
using the invariant measure (\ref{measure}) and restricting us to square
integrable functions, we obtain the physical states. The infinitesimal
generators of the symmetry group must have a hermitian character in 
order to be identified as observables of the system; in other words, we
must consider unitary representations.
Finally, the time evolution is given by a Schr\"odinger equation
$i\partial_{t}\Psi=H_\k \Psi$, where the Hamiltonian we are going to 
consider is essentially the Casimir (\ref{casex}),
$H_\k=\overline{C}_{\kappa}/2$ (it can be redefined up to additive or
multiplicative constants). Its explicit expression after substituting in
(\ref{casex}) the generators  by their associated vector fields (\ref
{campos-Oexb}) is
\bea\label{hamiltonian1} 
&&H_\k= -\frac{1}{2}\, \partial^{2}_{r} 
- \frac{\kappa}{2\sin^{2} {\skappa r}}\, \partial^{2}_{\theta}
+i\left(\kappa\,  \bb\,\vers r\, \frac{\cos \skappa
r}{\sin^{2}{\skappa r}} +\bb\right)\partial_{\theta} \nonumber\\ 
&&\qquad \qquad 
-\frac{\skappa}{2\tan{\skappa r}}\, \partial_{r} +
\frac{\kappa}{2\sin^{2}{\skappa r}}\, (\bb\vers r)^{2}. \label{quantsyst}
\eea

In general, the local representations are reducible, each irreducible
component is given by the Casimir equation
$\overline{C}_{\kappa}\Psi=\overline{c}_\k \Psi$. 
Whence, by construction, each eigenspace of $H_\k$ supports a unitary
irreducible representation (UIR) of $\sgroup$, since the eigenvalue
equation 
$H_\k\Psi_\k=\varepsilon_\k\Psi_\k$, 
$\varepsilon_\k=\overline{c}_\k/2$,
gives the irreducible subspaces of the local representation. The
description of our quantum  system will be complete if we compute the
spectrum, the degeneracy of the energy levels (given by the
aforementioned UIR) and a set of orthogonal eigenfunctions generating the
full Hilbert space of states.

\sect{Gauge potentials and minimal coupling\label{gaugepotential}}

In Section \ref{localrealigroup} we introduced the local realizations of 
$\group$ in a  direct operative way often used in the physics literature.
However, as we mentioned in Section 1, the natural framework for
the local realizations is the fiber bundle theory.  We shall analyze in this
section some properties obtained from  this more general viewpoint that
allow us to interpret physically (and geometrically) what is behind the 
Hamiltonian (\ref{hamiltonian1}) that we proposed in the preceding
section, and also it will help us to derive the minimal coupling rule for
interactions. For more details see Appendix D.

\subsect{Gauge invariant potentials}\label{gaugepotentials}

We can find a gauge invariant potential $A_\mu(x)$ under the action
(\ref{campos-Oexb}) of $\sgroup$. The local invariance condition of the
potential gives the following set of differential equations 
\begin{equation}\label{guaugepotencial}
X^\mu_j(x)\frac{\partial A_\nu(x)}{\partial x^\mu}
+ A_\mu(x)\frac{X_j^\mu(x)}{\partial x^\nu}
-i\frac{\partial W_j(x)}{\partial x^\nu} =0, 
\qquad \mu, \nu=1,2 ,\ \ \forall X_j\in \salgebra ,
\end{equation} 
where the fields $X_j(x)$, and the functions $W_j(x)$
of the local realization were defined in (\ref{locgen}) and 
(\ref{campos-Oexb}). It can be shown that this potential is the pull back of a
global invariant connection defined on a $U(1)$ principal bundle whose
base space is $\sphere$. 

The solutions to equation (\ref{guaugepotencial}),
taking coordinates $x^1=r$ and $x^2=\theta$, are  
\begin{equation}\label{potential}
A_{r}=0,\qquad  A_{\theta}= \bb\, \vers r.
\end{equation}
Such a solution is differentiable in a chart covering $\sphere$, except for
the south pole (as it was foreseeable, since the local realization was
smooth there).  This is the appropriate chart for our contraction  around 
the north pole. As usual, we can define the covariant derivatives by 
\begin{equation}\label{derivadacovariante}
D_{r}=-i\partial_{r}-A_{r},\qquad D_{\theta}
=-i\partial_{\theta}-A_{\theta} .
\end{equation}
Thus, the component of the invariant curvature form is
\begin{equation}\label{mag}
B_{r\theta}=- i\, [D_r,D_{\theta}] 
= \frac{\sin(\sqrt{\k} r)}{\sqrt{\k}} \,  \bb ,
\end{equation} 
which corresponds to a magnetic field normal to $\sphere$ whose intensity
is given by $\bb$ (recall the invariant measure (\ref{measure})). Remark that
if $\k \neq 0$, then $2\bb/\k \in \Z$, which in the case $\k=1$ is just the
Dirac monopole quantization \cite{dirac}. If we want that this intensity be
conserved along the limiting process $\k
\to 0$ we must take $\bb(\k) =
\bb_0$, i.e., a constant independent of $\k$. With this choice the potential
(\ref{potential}) has a well defined limit: $A_{r}=0, A_{\theta}=\bb_0\,
r^2/2$.

\subsect{Minimal coupling interactions}

Now we shall see how the minimal coupling rule can be introduced using 
arguments based on the symmetry algebra.

Let $\{X_{i}=X^{\mu}_{i}(x)\partial_{\mu}\}$ be the vector field 
realization (\ref{fields0}) on the pseudosphere $\sphere$ of the Lie algebra
basis of
$\algebra$, and let us consider the new set of generators
$X^{*}_{i}=X^{\mu}_{i}(x)D_{\mu}$, with $D_{\mu}$ the covariant
derivative (\ref{derivadacovariante}). As we said above, the Casimir
operator  $\overline{C}_\k({\overline X}_{i},B)$ (\ref{hamiltonian1}) of
$\overline{so}_{\k}(3)$ was identified (up to the factor $1/2$) with the
Hamiltonian of our  quantum system. Now, according to expression (\ref{ec})
of Appendix D, this operator can be obtained from the Casimir $C_{\k}(X_i)$
of $\algebra$  substituting the fields
$X_i$ by $X^{*}_{i}$: $\overline{C}_\k({\overline X}_{i},B)= C_{\k}(X^*_i)$.
Making use of this property we can rewrite the Hamiltonian
(\ref{hamiltonian1}) in terms of the vector fields $X^{*}_{i}$ as 
\bea\label{hamiltonian2} 
H_{\k}&=&
-\, \frac12\, D^{2}_{r}-\frac{\k}{\sin^{2}{\skappa r}}\, D^{2}_{\theta}
-\frac12\, \frac{\skappa}{\tan{\skappa r}}\, D_{r}\\ 
& =&-\,
\frac12\, \partial^{2}_{r} + \frac12\, \frac{\kappa}{\sin^{2}{\skappa r}}
\left(-i\, \partial_{\theta}-  \bb\vers r \right)^{2}
-\frac12\, \frac{\skappa}{\tan{\skappa r}}\, \partial_{r}. \nonumber 
\eea
The advantage of (\ref{hamiltonian2}) is that it makes explicit the
minimal coupling rule since it is the Hamiltonian of a free system on 
$\sphere$ where the derivatives have been replaced by covariant
derivatives. Therefore, (\ref{hamiltonian2}) describes the interaction of
a quantum system with an external magnetic field (\ref{mag}) normal to
the surface $\sphere$ given by the electromagnetic potential
(\ref{potential}).

The limit  $\k\to 0$ of the time-independent Schr\"odinger equation  
$H_\k \Psi_\k=\varepsilon_\k \Psi_\k$ is 
\begin{equation}
- \frac12\, \partial^{2}_{r}\Psi(r,\theta) + \frac{1}{2\,r^2}\, 
\left(-i\, \partial_{\theta}- \bb\, \frac{r^2}2\right)^{2}\Psi(r,\theta) -
 \frac{1}{2\,r}\, \partial_{r}\Psi(r,\theta)
=\varepsilon_0 \Psi(r,\theta).
\end{equation}
This is the eigenvalue equation in polar coordinates of a charged particle 
in the Euclidean plane under the action of a constant magnetic field of
intensity proportional to $|\bb |$ perpendicular to the plane, that is, the planar Landau
system \cite{landau}.

\sect{Eigenvalues  and eigenfunctions of Landau Hamiltonians
\label{uirso3kappa}}

\noindent
As a first step to get  the spectrum and eigenfunctions of the Hamiltonian
(\ref{hamiltonian2}) we will compute the  unitary irreducible
representations (UIRs) of $\sgroup$; afterwards, we will  analyze  their
relationship  with the  local representations determined in Section 
\ref{localrealigroup}.

\subsect{Elementary Landau quantum systems}\label{elementarylandauquantumsystems}

A basis of an UIR of $\sgroup$ is completely characterized by the 
eigenvalues and eigenvectors of three mutually commuting operators:
$H_{\k}\  (=\frac12\, {\overline C}_{\k})$, ${B}$ and $\overline{J}_{12}$. 
Let us denote by $\ketm$ an eigenvector of these operators, i.e.,
$$
H_{\k}\ketm=\vara \ketm,\  \overline{J}_{12}\ketm=m \ketm,
\  {B}\ketm=-\bb\ketm.
$$
The UIRs of the universal covering of $\sgroup$ can be obtained from 
those of $\salgebra$. We look for expressions valid for any $\k$-value
and at the same time for bounded representations. However, for $\k<0$ the
group is noncompact and there are other unitary representations (the
principal and supplementary series) not considered here.

Let $\{J^+,J^-, J, B\}$ be a ``Cartan'' basis of $\salgebra$,  where
${J}^{\pm} = \frac1{\sqrt2}(\overline{J}_{01} \pm i \overline{J}_{02})$ 
and  $J=\overline{J}_{12}$.
The non-vanishing
 Lie brackets are now
\begin{equation}
[{J},{J}^{\pm}] = \pm  {J}^{\pm},\qquad 
[{J}^+,{J}^-] = \k {J} + {B}.
\end{equation}
Since the representation must be unitary, then the generators must satisfy
the hermitian relations
$({J}^\pm)^{\dagger} =  {J}^\mp,\ {J}^{\dagger} = J,\ {B}^{\dagger} = {B}$. The second
order Casimir in the new basis reads 
\begin{equation}\begin{array}{ll}
\label{casimirnuevabase}
{\overline C}_{\kappa}&=2 J^{+} J^{-} + \k (J^{2} -J) +2{B}{J} -B\\[0.2cm]  
&=2 J^{-} J^{+} +\k (J^{2}+J) +2{B}{J} +B.
\end{array}
\end{equation}

There are two families of such bounded UIRs of $\sgroup$ characterized as
follows. One of them is given by a lowest negative weight, 
$-l\ (l\in \Z^+)$, such that ${J}^{-}\ketw =0$, and the other one by a
highest positive weight, $l\in \Z^+$, verifying ${J}^{+}\ketW =0$.

For the first family of UIRs the action of the generators $J^{\pm}$ on the
states $\ketm$ can be written as  
\be\label{uir}
\begin{array}{ll}
{J}^+\ketm &= \sqrt{ (l+m+1)(2\bb +{\k}(l-m))/2}\, \ketmp1 ,\\[.3cm]
{J}^- \ketm &= \sqrt{(l+m)(2\bb +{\k}(l-m+1))/2}\, \ketmm1 ,
\end{array}
\ee
with the restriction 
\be
(l+m)(2\bb +{\k}(l-m+1))\geq 0.
\ee 
This bounded representation is determined by the eigenvalue of the Casimir
(\ref{casimirnuevabase}), labeled by the integer $l$, 
\begin{equation}\label{eigenv}
\varaa ={\k}l(l+1)/2 +\bb(l+1/2).
\end{equation}

The  features of the UIRs of the first family depending on the particular 
values of $\k$ can be summarized as follows:

\begin{itemize}

\item
 $\k > 0$ :  $l \in \Z^{\geq 0}, \  2\bb/\k\in \Z$ such that 
$ -l< \bb/\k $.  The representation has dimension $2(l+\bb/\k)+1$
with carrier space generated by the set of eigenvectors
\be
\{\ketma\}^{l+2\bb/\k}_{m=-l}.
\label{k1}
\ee
Therefore, there are an infinite number of levels, each one finitely
degenerated.
 
\item
 $\k <0$ :  $l \in \Z^{\geq 0}$, $2\bb/\k\in \Z$ and 
$l <\bb/|\k|$.  We obtain an infinite-dimensional representation
with support space spanned by 
\be 
\{\ketma\}^{\infty}_{m=-l}.
\label{k2}
\ee
There is a finite number of discrete energy levels, $0\leq  l < \bb/|\k|$,
each one infinitely degenerated.

\item
 $\k=0$ : $l \in \Z^{\geq 0}$ and $\bb >0$. The representation is 
infinite-dimensional with a basis of the support space given by
\be
\{\ketma\}^{\infty}_{m=-l}.
\label{k3}
\ee
In this case we have infinite discrete energy levels infinitely degenerated.

\end{itemize}
In order to take in consideration the three cases together we will assume that 
$\bb >0$. In all of them $\varaa$ is given by (\ref{eigenv}).

For the second family of UIRs we have:
\be\label{uir2}
\begin{array}{ll}
{J}^+\ketm &= \sqrt{(l-m)(-2\bb +{\k}(l+m+1))/2}\,
\ketmp1  ,\\[.3cm]
{J}^- \ketm &=\sqrt{ (l-m-1)(-2\bb +{\k}(l+m))/2}\, \ketmm1 ,
\end{array}
\ee
with the restriction 
\be
(l-m)(-2\bb +{\k}(l+m+1))\geq 0.
\ee 
A bounded representation is determined by the eigenvalue of the Casimir
(\ref{casimirnuevabase}), labeled by the positive integer $l$,
\be\label{eigenv2} 
\varaa ={\k}l(l+1)/2 -\bb(l+1/2).
\ee
Similarly to the first family we have the three following cases according
with the values of $\k$:

\begin{itemize}

\item
$\k > 0$ :  $l \in \Z^{\geq 0}, \  2\bb/\k\in \Z$,  such that 
$ l> \bb/\k $. Basis: $\{\ketma\}_{-l+2\bb/\k}^{m=l}$.

\item
$\k <0$ :  $l \in \Z^{\geq 0},\  2\bb/\k\in \Z$, $-\bb > l |\k|$. Basis:
$\{\ketma\}_{-\infty}^{m=l}$.

\item
$\k=0$ : $l \in \Z^{\geq 0}$, $\bb < 0$.
Basis: $\{\ketma\}_{-\infty}^{m=l}$.

\end{itemize}
Now it is appropriate to consider $\bb< 0$ for the three cases, 
while $\varaa$ is given by (\ref{eigenv2}).
The same comments about dimensionality  and degeneration of the Landau
levels made for the other family of UIRs are also valid in this case.

 At this point we can check that the contraction process ($\k \to 0$)  
works correctly for the UIRs
defined above. For instance, the finite-dimensional representations of 
$\overline{SO}(3)$  ($\k >0$)
contract to infinite-dimensional ones of the two-dimensional Euclidean group
provided that $\k=2 |\bb|/n \to 0, \ n\in \N$, when $n\to\infty$ (for more
details see \cite{arratia97}). 

It is worth to remark that the energy eigenvalues 
(\ref{eigenv}) and (\ref{eigenv2}) include two terms. The first of them is
quadratic in $l$ and has a geometric character through the curvature $\k$ of
the configuration space. The second term, linear in $l$, is the only one that will
remain in the planar limit and has a dynamic character by means of $\bb$ 
that was interpreted as a magnetic field.

\subsect{A complete set of eigenfunctions}\label{completeset}

Once the above UIRs of $\sgroup$ has been characterized we have to check
whether they are realizable as irreducible components of the local
representations given in Section 3.

Recall that we must restrict to differentiable wavefunctions around the
north pole. So, we can write  
$\Psi_{l,m}^{\bb,\k}(r,\theta) = e^{im\theta}R_{l,m}^{\bb,\k}(r)$, $m\in \Z$ 
as the wavefunction associated to the basis element
$\ketm$, i.e.,  $\langle r,\theta\ketm=\Psi_{l,m}^{\bb,\k}(r,\theta)$, of a
lowest weight representation with Casimir eigenvalue (\ref{eigenv}) given 
by ${\overline c}_{\k} =\varaa/2$.  From (\ref{campos-Oexb}) the
local expression of the up and down operators  take the form
\be
J^{\pm} =  i\, e^{\pm i\,\theta}\left( - \partial_r \, {\mp} \,
\frac{i\sqrt{\k}}{\tan\sqrt{\k}r}\partial_{\theta} \, {\pm} \,
{\bb}\,{\rm ver}_{\k}(r) \frac{\sqrt{\k}}{\sin\sqrt{\k}r} \right).
\ee
For each eigenvalue, $m$, of $J \equiv J_{12}$ we can compute the radial
eigenfunctions $R_{l,m}^{\bb,\k}(r)$ quite easily. So, the
fundamental wavefunction corresponding to $|\vara,\bb,{-}l\rangle$ is
determined by the equation 
\be\label{eigenr}
\left( - \frac{d}{dr} + 
\frac{l\, \sqrt{\k}}{\tan\sqrt{\k}r} - 
{\bb}\,{\rm ver}_{\k}(r) \frac{\sqrt{\k}}{\sin\sqrt{\k}r} \right)\, 
R_{l,-l}^{\bb,\k}(r) = 0,
\ee
whose solution (up to normalization) is
\be\label{rr}
R_{l,-l}^{\bb,\k}(r) = \left(\frac{\sin\skappa
r}{\skappa}\right)^{l}\left(1+\tan^{2}\frac{\skappa
r}{2}\right)^{-\bb/\k}. 
\ee
The complete function $\Psi_{l,-l}^{\bb,\k}(r,\theta) =
e^{-i\, l \theta}R_{l,-l}^{\bb,\k}(r)$ is square integrable on the `sphere'
$S_{\k}^2$  with respect to the invariant measure (\ref{measure}) if 
$0\leq l <\bb/|\k| - 1/2$ for $\bb >0$. From 
this eigenfunction and using the raising operator
$J^+$  we can find all the remaining basis eigenfunctions generating the 
whole $\varaa$-eigenspace:
\be \label{psi}
\Psi_{l,-l+n}^{\bb,\k}(r,\theta) \propto (J^+)^n \, 
\Psi_{l,-l}^{\bb,\k}(r,\theta).
\ee
These eigenfunctions are also square integrable provided the requirements 
(\ref{k1})--(\ref{k2}) of the previous subsection  are fulfilled besides $l <\bb/|\k| - 1/2$. In this
way we have completed the search of the spectrum and eigenfunctions of 
the Hamiltonian (\ref{hamiltonian2}) for any value of $\k$. 

Remark that for $\k = 0$ the Landau energy levels (\ref{eigenv}) are
linear in $l$, $\varaao= 2\bb\, (l +1/2)$. The fundamental state inside the
$\varaao$-eigenspace is obtained by taking the limit $\k\to 0$ of 
(\ref{rr}),
\be
R_{l,m}^{\bb,0}(r) = r^{l} e^{-\bb\, r^2/4}.
\ee
This can be used to derive the rest of the infinite basis eigenfunctions 
with the help of the shift operators
$$
J^{\pm}= i\, e^{\pm i\,\theta}\left( - \partial_r  \mp
\frac{i}{r}\partial_{\theta}  \pm  \frac{{\bb}\, r}{2} \right).
$$ 

The results for the second family of UIRs are quite similar taking into
account obvious sign changes in $m$ and $\bb$  (recall in this respect that
now $\bb$ is negative). For instance,  the fundamental state is 
$\Psi_{l,l}^{\bb,\k}(r,\theta) = e^{i\, l \theta}R_{l,l}^{\bb,\k}(r)$, 
where 
\be
\label{rrr}
R_{l,l}^{\bb,\k}(r) = \left(\frac{\sin\skappa r}{\skappa}\right)^{l}
\left(1+\tan^{2}\frac{\skappa r}{2}\right)^{\bb/\k}.
\ee

\subsect{Lowest Landau level}\label{lowestlandau}

A case with an special interest  is when $l=0$. Once fixed the geometry
(i.e., $\k$) and the external field ($\bb$) this value corresponds to  the lowest
energy Landau level. From (\ref{psi}) it is easy to show that the (radial
component of the) eigenfunctions are simply given by
\be \label{lowest}
R_{0,m}^{\bb,\k}(r) =
N(\bb,\k,m)\, 
\left(\cos \skappa r/2 \right)^{2\bb/\k -m}
\left(\frac{\sin\skappa r/2}{\skappa}\right)^{m}\, ,
\ee
with the normalizing coefficient
\be
N(\bb,\k,m) = 
\left(
\frac{2(2\bb+\k)\dots(2\bb +\k -m\k)}{\Gamma(m+1)}
\right)^{1/2}.
\ee
If we further select $\k=0$ the above formulae reduce to the
lowest level of the planar Landau system, whose eigenfunctions are
\be
R_{0,m}^{\bb,0}(r) =
N(\bb,0,m)\, (r/2)^m\, e^{-\bb r^2/4}\, ,
\ee
with $ N(\bb,0,m) = (2^m \bb^{m+1}/\Gamma(m+1))^{1/2}$.

The state density  (degeneracy of the $l$-th level/area) for the spherical case is 
\be
\frac{2l+1+2|\bb|\k}{4\pi /k}=\frac{|\bb|}{2\pi }+\frac{\k(2l+1)}{4\pi}.
\ee
In the Euclidean limit ($\k=0$) the state density is $|\bb|/2\pi$ for any $l$ in
agreement with the Landau result \cite{landau}.  In the hyperbolic case the
state density is also  $|\bb|/2\pi$.

\subsect{Eigenfunctions in terms of hypergeometric functions}\label{eigenfunctions}

The components $R_{l,m}^{\bb,\k}(r)$ of the basis wavefunctions can be 
written in terms of hypergeometric functions. In order to achieve this,
we start from the  Schr\"odinger equation for the eigenvalue $\varaa$
(\ref{eigenv}) 
\be\label{hyper1}
\left(\frac{d^{2}}{dr^{2}}
+ \frac{\skappa}{\tan\skappa r}\frac{d}{dr}
- \frac{\kappa}{\sin^{2}\skappa r}(m-\bb\vers r)^{2}
+ 2 \varaa\right) R_{l,m}^{\bb,\k}(r)=0.
\ee
If we change to the new variable $x=A\vers r$ and factorize the
wavefunction as  
\be \label{fase}
R_{l,m}^{\bb,\k}(r)=A^{m/2}\vers^{m/2} r\, 
(1-\frac{\kappa}{2}\vers r)^{\bb/\kappa -m/2}\phi_{l,m}^{\bb,\k}(r),
\qquad 
\Phi(x) = \phi(r(x)),
\ee
we can  rewrite this equation as an hypergeometric-like equation
\be\label{hyper2}
x(1-\frac{\kappa}{2A}\, x)\frac{d^{2}\Phi_{l,m}^{\bb,\k}(x)}{dx^{2}}
+ (1+m-\frac{\bb+\kappa}{A}\, x)\frac{d\Phi_{l,m}^{\bb,\k}(x)}{dx}
+ \frac{2\varaa-\bb}{2A}\, \Phi_{l,m}^{\bb,\k}(x)=0 .
\ee
Remark that the factor function in (\ref{fase}) coincides with the Landau
eigenfunctions of the  lowest level (\ref{lowest}). It is also worth to note
that equation (\ref{hyper2}) is well behaved when $\k \to 0$ giving rise to a
confluent hypergeometric equation. 

We shall analyze in detail eq.\  (\ref{hyper2}) according to the values of $\k$:

\begin{itemize}

\item[i)]
$\kappa \neq 0$. Choosing $A=\kappa /2$, eq.\ (\ref{hyper2}) turns into
the hypergeometric expression
\be\label{hyper3}
x(1-x)\frac{d^{2}\Phi_{l,m}^{\bb,\k}(x)}{dx^{2}} 
+ (1+m-2({\bb}/{\kappa}+1)x)\frac{d\Phi_{l,m}^{\bb,\k}(x)}{dx}
+ \frac{2\varaa -\bb}{\kappa}\,\Phi_{l,m}^{\bb,\k}(x)=0,
\end{equation}
and the solutions we are looking for are given in terms of the 
hypergeometric function
$\Phi_{l,m}^{\bb,\k}(x)=F(-l,l+1+2\bb/\kappa,m+1,x)$. However, in order to
avoid problems when $1+m\leq 0$ we can consider \cite{olver}  
$$
\F(-l,l+1+2\bb/\kappa,m+1,x)=\frac {F(-l,l+1+2\bb/\kappa,m+1,x)}
{\Gamma (m+1)} ,
$$
which is also solution of (\ref{hyper3}).
The complete expression of the local eigenfunctions is
\bea\label{completafuncion}
&& \Psi_{l,m}^{\bb,\k}(r,\theta)
= c_{lm}(\kappa) \frac{\kappa^{m/2}}{2^{m/2}}\, e^{i\,m\theta}\,
\vers^{m/2} r\, (1{-}\frac{\kappa}{2}\vers r)^{\bb/\kappa {-}m/2}
\nonumber\\[0.3cm] 
&& \hskip4cm \times  \F(-l,l{+}1{+}2\bb/\kappa,m{+}1,
\frac{\kappa}{2}\vers r),
\eea
where the factor $c_{lm}(\kappa)$ is a normalization constant. With the 
measure (\ref{measure}) its value is
\be\label{coeflm}
c_{lm}(\kappa)=\sqrt{\frac{\kappa\Gamma (l+1+2\bb/\kappa)
\Gamma (2l+2+2\bb/\kappa)\Gamma(l+m+1)}{4\pi\Gamma (2l+1+2\bb/\kappa)
\Gamma (l-m+1+2\bb/\kappa)}} \, .
\ee

Since the above hypergeometric functions can also be expressed in terms of 
the Jacobi functions of first kind, 
${\cal P}^{(m,2\bb/\kappa-m)}_{l}(\cos\skappa r)$, we can rewrite
(\ref{completafuncion}) as  
\bea
&& \Psi_{l,m}^{\bb,\k}
(r,\theta)=c_{lm}(\kappa)\frac{\kappa^{m/2}}{2^{m/2}}
\, e^{i\,m\theta}\,
\vers^{m/2}  r (1-\frac{\kappa}{2}\vers r)^{\bb/\kappa -m/2}
\nonumber\\[0.3cm]  
&& \hskip3cm \times 
\frac{l!}{\Gamma (l+m+1)}  P_l^{(m,2{\bb}/{\kappa} -m)}(\cos\skappa r).
\eea

\item[ii)]
$\kappa = 0$.  Let us assume $\bb \neq 0$ and choose $A=\bb$, then 
eq.\  (\ref{hyper2}) comes into one of the confluent hypergeometric
class 
\be\label{hyper4}
x\, \frac{d^{2}\Phi_{l,m}^{\bb,\k}(x)}{dx^{2}}
+ (1+m-x)\, \frac{d\Phi_{l,m}^{\bb,\k}(x)}{dx}+l\,
\Phi_{l,m}^{\bb,\k}(x)=0.
\ee
The appropriate solutions are expressed by means of the confluent 
hypergeometric function  \cite{olver} 
$M(-l,m+1,x)$, or $\M(-l,m+1,x)=M(-l,m+1,x)/\Gamma (m+1)$.
So,  we obtain
\be
\Psi_{l,m}^{\bb,0}(r,\theta)=c_{lm}(0)\, 
\frac{\bb ^{m/2}}{2^{m/2}}\, e^{i\, m\theta}\,r^{m}e^{-\bb r^{2}/4} 
\M(-l,m+1,\bb r^{2}/2),
\ee
where the normalization constant $c_{lm}(0)$ is given by
\be
c_{lm}(0) = \sqrt{\frac{\bb\, \Gamma (l+m+1)}{2\pi\, l!}} \, .
\ee
In terms of Laguerre polynomials, $L^{m}_{l}(\bb r^{2}/2)$, the
solutions $\Psi_{l,m}^{\bb,0}(r,\theta)$ can be rewritten as 
\begin{equation}
\Psi_{l,m}^{\bb,0}(r,\theta)=c_{lm}(0)\, 
\frac{\bb ^{m/2}}{2^{m/2}}\, e^{i\, m\theta}\, r^{m}e^{-\bb
r^{2}/4}\frac{l!}{\Gamma (l+m+1)}L^{m}_{l}(\bb r^{2}/2)\, .
\end{equation}

\end{itemize}
It can be checked that the following limits when $\kappa\to 0$ hold:
\bea
&&\lim_{\kappa\to 0}c_{lm}(\kappa)\frac{\kappa^{m/2}}{2^{m/2}}
= c_{lm}(0)\bb ^{m/2}, \nonumber\\
&&\lim_{\kappa\to 0}
\vers^{m/2}  r \, (1-\frac{\kappa}{2}\vers r)^{\bb/\kappa -m/2}
= \frac{1}{2^{m/2}}\, r^{m}e^{-\bb r^{2}/4},\\ 
&&\lim_{\kappa\to 0} F(-l,l+1+2\bb/\kappa,m+1,\frac{\kappa}{2}\vers r)
= M(-l,m+1,\bb r^{2}/2). \nonumber
\eea
These limits prove that there exists a well defined contraction process 
for the local UIRs wavefunctions given by 
\be
\lim_{\kappa\to 0}\Psi_{l,m}^{\bb,\k}(r,\theta)
=\Psi_{l,m}^{\bb,0}(r,\theta).
\ee

The second family of UIR's admits a similar treatment. 
Now, factorizing  
$$
R_{l,m}^{\bb,\k}(r)=A^{-m/2}\vers^{-m/2} r\, 
(1-\frac{\kappa}{2}\vers r)^{-\bb/\kappa +m/2}\phi_{l,m}^{\bb,\k}(r)
$$ 
and  performing the variable change $x=A\vers r$ we obtain 
\bea
&& \Psi_{l,m}^{\bb , \k}(r,\theta)
= c_{l,-m}(\kappa)\frac{2^{m/2}}{\kappa^{m/2}}
\, e^{im\theta}\, \vers^{-m/2} r
(1-\frac{\kappa}{2}\vers r)^{-\bb/\kappa +m/2} \nonumber\\[0.3cm]  
&& \hskip3cm \times 
\frac{l!}{\Gamma (l-m+1)} 
P_l^{(-m,-2\frac{\bb}{\kappa} +m)}(\cos\skappa r) ,
\eea
with $c_{l,-m}$ given by (\ref{coeflm}). This function is a solution of 
eq.\  (\ref{hyper3})
where $m$ and $\bb$ have been replaced by $-m$ and $-\bb$, respectively.

\sect{Horocyclic coordinates and variable separation\label{horocycliccoordinates}}

In this section we shall perform a coordinate separation of  the hyperbolic Landau quantum system
by means of horocyclic coordinates \cite{olevsky}. The same question, but at the classical level
in the complex plane, was addressed  in Ref.~\cite{duru,comtet}.  Nowadays  
the hyperbolic Landau classical problem continues to be  a matter of study from different points
of view (see for instance \cite{almorox,tejero} and references therein).

There are two remarks worth to mention on this subject. The first one is that in the convention of
Miller  \cite{miller} the variable separation in the quantum case is a example of $R$--separability,
i.e., given an equation $E\psi=0$ the $R$--separable solutions  are, in fact, standard separable
solutions of an equivalent equation $E'\phi=0$ with $E'=R^{-1}E R$ and  $\psi=R\phi$. The second
remark is that horocyclic coordinates under the contraction $\k \to 0$ turn into cartesian
coordinates in the plane.

The horocyclic coordinates
$(a,b)\in \R^2$ are associated to the action of the  generators
$J_{01}$ and
$J_{02}+\sqrt{-\k}J_{12}$ of $\group$  over the point  $x_0=(1,0,0)$ of $S_{\k}^2$ as follows (in
this section  $\k < 0$) 
\be
(x^0,x^1,x^2)^T=  e^{- i a J_{01}} e^{- i b (J_{02}+\sqrt{-\k}J_{12})} (1,0,0)^T ,
\ee
where the superindex $T$ means matrix transposition and  the matrix representation of the 
generators $J_{..}$ is given in expression (\ref{matrixrepresentation}).
The explicit expression of this coordinate system is
\be\label{horociclicas}
\begin{array}{lll}
x^0 &=& \cosh \sqrt{-\k}a -\k\ \frac {b^2}{2}\ e^{\sqrt{-\k} a} ,\\[0.2cm]
x^1 &=&  \frac {\sinh (\sqrt{-\k}a)}{\sqrt{-\k}} +\k \ \frac {b^2}{2}\ e^{\sqrt{-\k}a} ,\\[0.2cm]
x^2 &=&   b \ e^{\sqrt{-\k}a} .
\end{array}\ee
 In the limit $\k \to 0$ we recover the cartesian coordinates of the plane as we mentioned
above. These horocyclic coordinates are of ``subgroup type'' like  the  polar geodesic ones used
in previous sections. While   the former   corresponds to the reduction
$O(2,1)\supset T$, where $T$ is the subgroup generated by $J_{02}+\sqrt{-\k}J_{12}$, the last one 
is related to $O(2,1)\supset O(2)$ (for more details see \cite{boyer} and references therein).

Using  horocyclic coordinates the following time-independent Schr\"odinger equation of
the hyperbolic Landau systems  holds
\be\label{landauecuacionab}
\left[\frac{1}{\k}(\frac{\partial}{\partial a}- i V_a )^2
-\frac{1}{\sqrt{-\k}}(\frac{\partial}{\partial a}- i V_a) +\frac{e^{- 2 \sqrt{-\k}
a}}{\k}(\frac{\partial}{\partial b}- i V_b)^2\right]\Phi(a,b)=-\frac{E}{\k}\Phi(a,b)
\ee
where 
\be\label{potencialesab}
V_a=\frac{2\  \k\ \bb \  b}{2- \k b^2 e^{\sqrt{-\k} a} + 2 \cosh{\sqrt{-\k} a}},\qquad
V_b=\frac{\bb (-1+ e^{-2\sqrt{\k} a}(1-\k b^2))}{2- \k b^2 e^{\sqrt{-\k} a} + 2 \cosh{\sqrt{-\k} a}}
\ee
are the electromagnetic potential components in these coordinates (see subsection
\ref{gaugepotentials}).

Taking under consideration that the wave function has the form
\be\label{phase}
\Phi(a,b)=\exp{\left[{\frac{\bb}{\sqrt{-\k}}}\left(\sqrt{-\k}\; b 
-2 \arctan \frac{\sqrt{-\k}\ b \ e^{\sqrt{-\k} a}}{1+e^{\sqrt{-\k} a}}\right)\right]}\psi(a)\phi(b)
\ee
and 
\be\label{funcionb}
 \phi(b)=e^{i \lambda b} ,\qquad \lambda \in \R, 
\ee
we obtain, after rescaling multiplying by $\k$, a new differential equation only in the variable $a$,
i.e., coordinates
$a$ and $b$ allow a variable separation with $\lambda$ as the separation constant 
\be\label{landauecuaciona1}
\left[-\frac{\partial ^2}{\partial a^2} -{\sqrt{-\k}}\frac{\partial}{\partial a} -
\frac{1}{\k} e^{-2\sqrt{-\k}a} (\beta (e^{\sqrt{-\k} a} -1)-\sqrt{-\k} \lambda)^2 -{E}
\right]\psi(a)=0.
\ee
Note that the differential operator $\overline{J}_{02}+\sqrt{-\k}\; \overline{J}_{12}$
(see expression \ref{campos-Oexb}) is straightened out to the form 
$- i \partial_b +  f(a,b)$, where
$$
f(a,b)= \frac{2\sinh(\sqrt{-\k} a)+\k b^2 e^{\sqrt{-\k} a}}{\sqrt{-\k}(2-\k e^{\sqrt{-\k}
a}+2\cosh(\sqrt{-\k} a))}
$$ 
corresponds to the term $W_{02}(x)$ of $\overline{J}_{02}$ in (\ref{campos-Oexb}). In order to
eliminate this function we introduce the phase (\ref{phase}) of  $\Phi(a,b)$, which  performs the
$R$-separation and it is defined by $\exp (\int f(a,b) db)$. The operator $\overline{J}_{01}$
given by (\ref{campos-Oexb}) becomes $\overline{J}_{01}=-i \partial_a + \bb b$. 

In the limit $\k \to 0$ of equation (\ref{landauecuaciona1}) we recover the harmonic oscillator
Schr\"odinger equation of unit mass, frequency $\omega=|\bb|$, energy $E/2$ and origin
$a=\lambda/\bb$. 

Equation (\ref{landauecuaciona1}) can be set into the standard expression
\be\label{morsecuaciona1}
\left[-\frac{\partial ^2}{\partial a^2}  -\left( E' -
\frac{\bb ^2}{\k}(- e^{-2\sqrt{-\k}a} +2 e^{-\sqrt{-\k} a} )\right) 
\right]\psi(a)=0 ,
\ee
with 
\be\label{morsenergy}
E'=E- \frac{\bb ^2}{\sqrt{-\k}} + \frac{\k}{4}\ ,
\ee 
by means of the following transformations:
\begin{itemize}

\item coordinate translation $a\to a -\alpha/\sqrt{-\k}$, with 
$e^\alpha=(\bb +\sqrt{-\k} \lambda)^2$
and ${\rm sign}(\bb)={\rm sign}(\lambda)$,

\item $\psi(a) \to e^{-\sqrt{-\k}/2} \psi(a)$,

\item new coordinate translation $a\to a -\gamma/\sqrt{-\k}$, where $|\bb|=e^{\gamma}$.
\end{itemize}

\noindent
Equation (\ref{morsecuaciona1}) corresponds to the Schr\"odinger equation of a particle of
unit mass moving in a Morse potential $\frac{\bb ^2}{2\k}(-e^{-2\sqrt{-\k}a} + 2 e^{-\sqrt{-\k}
a})$   \cite{landau}. 

As it is well know the energy spectrum of the Morse potential has two parts: one discrete ($E'< 0$)
and other continuous ($E' \geq 0$). Hence, we have from (\ref{morsenergy}) that the energy for the
continuous spectrum of the  Landau problem corresponds to
\be
E\geq \frac{\bb ^2}{\sqrt{-\k}} - \frac{\k}{4}
\ee

In order to obtain the discrete spectrum we proceed as follows: the change of variable
\be
\xi=-\frac{2\sqrt{\bb ^2}}{\k} e^{-\sqrt{-\k} a}
\ee
and the  factorization 
\be
\psi(a)=e^{-\xi /2} \xi^s f(\xi)
\ee
gives the equation
\be\label{morsecuacion}
\xi f''(\xi)  + (2s +1 - \xi) f'(\xi) + l f(\xi) =0 ,
\ee
where
\be
s=\frac{\sqrt{-E'}}{\sqrt{-\k}} , \qquad  l=-\frac{\sqrt{\bb^2}}{\k} -(s+\frac 12) .
\ee
The confluent hypergeometric function
\be
f(\xi)= M(-l, 2s +1, \xi) \qquad l\in \Z ^{\geq 0}.
\ee
is solution of equation (\ref{morsecuacion}).
The energy spectum is
\be\label{morsecuacionespectro}
{\cal E}_l=\frac {E_l}{2}=|\bb| (l+\frac 12) + \frac{\k}{2} l (l+1)
\ee
that agrees with expressions (\ref{eigenv}) and (\ref{eigenv2}). The number of Landau levels is
finite since
\be
0\leq l< \frac{|\bb|}{|\k|} - \frac{1}{2} .
\ee

\sect{Algebraic analysis of  Landau equations}\label{algebraicanalysis}

\subsect{Ladder operators for energy levels}

For the planar Landau systems, besides the (shift)
operators that act inside each energy level changing only the values of $m$,
there are also other type of (ladder) operators  connecting states of different
energies. We shall show in the following that the Landau systems with 
$\k \neq 0$ (thus, including the spherical and hyperbolic systems) also admit
ladder operators such that in the limit $\k\to 0$ come into those associated
to the planar case.

Let us multiply the eigenvalue equation (\ref{hyper1}) by the function
$({\sin^{2}\skappa r})/{\kappa}$. The resulting differential equation 
\be \label{abc}
{\cal E}_l\,R_{l,m}^{\bb,\k}(r)\equiv \left(\frac{\sin^{2}\skappa r}{\kappa}\,
\frac{d^{2}}{dr^{2}} {+ } 
\frac{\sin\skappa r \, \cos\skappa r}{\skappa }\frac{d}{dr} 
{-} (m{-}\bb\vers r)^{2} 
{+ } \frac{\varaa\, \sin^{2}\skappa r}{\kappa}\right)
R_{l,m}^{\bb,\k}(r)=0 
\ee
can be factorized as follows
\be \label{qq}
\left\{\left(\frac{\sin\skappa r}{\skappa}\, \frac{d}{dr}{+}\mu_l
\,\cos\skappa r
{+}\nu_l\right)
\left(\frac{\sin\skappa r}{\skappa}\, \frac{d}{dr}{-}\mu_l\,\cos\skappa r
{-}\nu_l\right) {+}
\delta_l\right\} R_{l,m}^{\bb,\k}(r)=0,
\ee
with
\be\begin{array}{l}\label{aa}
\mu_l = \frac{\bb}{\k} +l,\qquad \quad \nu_l =-\frac{\bb}{\k}
+\frac{\bb(m+l)}{\bb +\k  l} \\[0.3cm]
\delta_l = \frac{2\bb l(m+l)}{\bb +\k l} +\frac{\bb^2(m+l)^2}{(\bb + \k
l)^2} -m^2 + l^2 .
\end{array}\ee
So, the factorization of the second order differential operator ${\cal E}_l$ in
(\ref{abc}) can also be written schematically in the form
\be\label{pp}
{\cal E}_l = A^+_l A^-_l +\delta_l,
\ee
where the label $l$ corresponds to the energy value $\varaa$ in eq.\ 
(\ref{eigenv}) keeping fixed $\k$ and $m$. It can be checked (for instance,
through the symmetry change $l \to -l -2\bb/\k -1$), according to the
previous notation, that
\be
{\cal E}_{l-1} = A^-_l A^+_l +\delta_l= A^+_{l-1} A^-_{l-1} +\delta_{l-1} .
\ee
This means that the operator $A^-_l$ connects the eigenfunction 
space of eigenvalue $\varaa$ to that one corresponding to 
$\varaaa$, while $A^+_l$ acts in the opposite direction. In fact, when
$A^{\pm}_l$ do not spoil the normalization conditions, they will link
the radial eigenfunctions $R_{l,m}^{\bb,\k}(r)$ and $R_{l\pm 1,m}^{\bb,\k}(r)$,
up to a factor. By means of the operator set $\{A^{\pm}_l\}$ we
can define free-index operators $A^{\pm}$ (see ref.\ \cite{miller}) with 
commutation  rules
\be \label{b}
[A^-,A^+] = \Delta(L),
\ee
where the involved operators in (\ref{b}) when acting on $R_{l,m}^{\bb,\k}(r)$
must be  read in the form: 
\bea
&&[A^-,A^+]= A^-_{l{+}1}A^+_{l{+}1} -A^+_{l}A^-_{l}\nonumber\\ [0.2cm]
&&\Delta(L) = \delta_{l}-\delta_{l{+}1},
\label{c}
\eea 
being $L$ a diagonal operator, $L\,R_{l,m}^{\bb,\k}= l\,R_{l,m}^{\bb,\k}$.

In the limit $\k\to 0$ all the elements in (\ref{qq})--(\ref{aa}) are well
defined and  we recover the planar Landau ladder operators:
\bea
&&A^{\pm}_l \to  r \frac{d}{dr} \mp \frac{\bb}2 r^2 \pm (2l+m) \nonumber\\[0.2cm]
&&\delta_l \to 4l^2 +4lm.
\eea
Now, the so obtained free-index operators $\{A^+, A^-,L\}$, for $\k =0$, close
a Lie algebra isomorphic to $so(2,1)$. However, when $\k \neq 0$, as can be
seen from (\ref{pp}) and  (\ref{b}), these operators generate an associative algebra but not a
Lie algebra. 

There is a freedom in normalizing the operators $\{A^+, A^-\}$ of (\ref{c}), so
that if we change to the set $\{\tilde A^+_{l}=\sqrt{\bb +\k l}A^+_{l},
\tilde A^-_{l}=A^-_{l}\sqrt{\bb +\k l}\}$, now the new pair  
$\{\tilde A^+, \tilde A^-\}$ will satisfy cubic commutation relations. This is
the kind of algebra related to the isotropic oscillator in curved spaces
discussed in Refs.\  \cite{higgs}--\cite{natalie}. Such a connection seems very
suggestive since as it is known the Landau system in the plane is closely
related to the two-dimensional oscillator.

\subsect{Annihilation lines and solution sectors}

We shall study some consequences of the factorization (\ref{qq}) that can
help in computing the eigenfunctions of the Landau wave equation by a new
procedure. This section can be seen as an application of the refined
factorization method \cite{olmo96,refined}.

The main role of our discussion is played by the expression of $\delta_l$
in (\ref{aa}). Although apparently complicated, it is the responsible of the
spectrum `shape' of the Landau systems in a way that will be precised below.
Let us consider the solutions in $l$ of the equation 
\be
\delta_l = \frac{2\bb l(m+l)}{\bb +\k l} +\frac{\bb^2(m+l)^2}{(\bb + \k
l)^2} -m^2 + l^2 =0.
\ee
For these values, according to (\ref{qq}), the states $\psi^-_l$ annihilated by
$A^-_l$, $A^-_l\psi^-_l=0$, satisfy the equation (\ref{abc}) 
${\cal E}_l \psi^-_l=0$, in other words, they are solutions of the Landau
eigenequation. In the same way the reasoning goes through for the solutions
of $\delta_{l+1} = 0$. In this case the states $\psi^+_l$ such that 
$A^+_{l+1}\psi^+_l=0$ will be also solutions of (\ref{abc}).

Based on these considerations, once fixed $\k$, the solutions of 
$\delta_l =0$ ($\delta_{l+1} =0$) in the plane $(m,l)$ will be called
annihilation lines of $A^-$ ($A^+$). These lines provide immediate solutions
of the Landau systems, but it is still necessary to specify carefully which ones
are normalizable. In this case such states will constitute vacuum states
that can be used to build the whole spectrum by applying ladder operators.
It is also important to determine when the action of such operators will lead
us out of the normalizable sector.

The solutions to the equation $\delta_l =0$ are straight lines. For each of 
these lines we can build the operators $A^-_l$ according to
(\ref{qq})--(\ref{pp}) and find the cases where the states $\psi^-_l$ are
normalizable. The results are summarized below depending on the $\k$ values
(we have always assumed that $\bb> 0$):

\begin{itemize}
\item
($\k=0$)  
\be\label{1}
\begin{array}{lccc}
& &{\rm solutions}\qquad\qquad &\psi^-_l  {\rm \   \ normalizable \  if}\\[0.3cm]
& i)\quad &l=-m,\qquad\qquad &m\leq 0 \\[0.2cm]
&ii)\quad &l=0,\qquad\qquad  &m\geq 0
\end{array}
\ee
\item
($\k > 0$)    
\be\label{2}
\begin{array}{lccc}
& &{\rm solutions}\qquad\qquad &\psi^-_l  {\rm \   \ normalizable \  if}\\[0.3cm]
&i)\quad &l=-m,\qquad & m\leq 0  \\ [0.2cm]
&ii)\quad &l=0,\qquad & 0\leq m \leq 2\bb/\k  \\[0.2cm]
&iii)\quad &l=-2\bb/\k,\qquad & {\rm never} \\[0.2cm]
&iv)\quad & l=m - 2\bb/\k , \qquad & m\geq 2\bb/\k . 
\end{array}
\ee
Since the UIRs of $\group$ ($\k>0$), computed in subsection  
\ref{elementarylandauquantumsystems}, restrict the parameter values to $2\bb/\k \in \Z^+$, 
$l\in \Z^+$, 
$-l \leq m\leq 2\bb/\k$, we see that they coincide with those displayed in
the above table. In fact, we can check that the normalizable eigenfunctions
$\psi^-_l$ defined on the annihilation lines are the same than those
previously found in subsections \ref{completeset} and \ref{lowestlandau}. Therefore, the
shift and ladder operators are consistent in the sense that they act in the same space of
physical states.
\item
($\k < 0$)
\be \label{3}
\begin{array}{lccc}
& &{\rm solutions}\qquad\qquad &\psi^-_l  {\rm \   \ normalizable \  if}\\[0.3cm]
&i)\quad &l=-m,\qquad & \bb/\k +1/2 < m \leq 0  \\[0.2cm]
& ii)\quad &l=0,\qquad & 0\leq m  \\[0.2cm]
&iii)\quad &l=-2\bb/\k,\qquad & {\rm never} \\[0.2cm]
&iv)\quad &l=m - 2\bb/\k , \qquad & {\rm never}   
\end{array}
\ee
The same comments can be done with respect to this table: all the
restrictions and wavefunctions are consistent with the unitary
representations of $\group$ with $\k <0$.
\end{itemize}

The sector of physical eigenstates bounded by these lines are
depicted in Fig.~1. The parameters associated to the normalizable wave
functions that constitute a lattice inside such sectors are shown
schematically in  Fig.~ 2.

In order to look for the annihilation lines of the operator $A^+$ one can use
the symmetry $l \to -l -2\bb/\k -1$ to get:
\be
i') \  l=m-2\bb/\k -1,\quad
ii') \  l = - 2\bb/\k -1,\quad
iii') \  l = -1,\quad
iv') \  l = -m -1.
\ee
These lines can be used to give an equivalent description for the unitary
representations corresponding lo highest weight $l\in \Z^-$ quoted in subsection  
\ref{elementarylandauquantumsystems}, so that we shall not refer to them any longer.
 The graphs of physical
sectors, lines and states are symmetric, with respect to the $l$ axis, to the
cases derived from $A^-$.

\subsect{Moving states}

If we exclude the north pole of $\sphere$ (or both, if $\k > 0$), then the set
of local realizations of the Lie algebra $\algebra$ acting on differentiable
functions is bigger, since it is parametrized by two real labels $b$ and $\la$, as it
is shown in Appendix C. 

Let us concentrate on the case $\k \neq 0$. The $\theta$--component of the invariant gauge
potential is $A_{\theta}=  b/\k-\lambda\cos(\sqrt{\k}r)$, hence  changes in the parameter $b$
lead to the same field (see Appendix C and subsection
\ref{gaugepotentials}).  Also, we can perform a gauge transformation in the class of
differentiable functions changing $b/\k$ into
$b/\k + n$, $n\in \Z$.  Moreover if we mantain
the notation of previous sections $\la = \beta/\k$,  now the potentials can be rewritten as 
$A_{\theta}=  \beta/\k\, {\rm vers}_{\k}(r) + \rho$ with $\rho = b/\k - \la +n$. 
Therefore, the classes of gauge equivalent potentials can be characterized by
a real parameter $0\leq \ro <1$ \cite{asorey}.

We can change the point of view and leave one potential fixed for each field (choosing for instance
$\ro =0$) but defining different classes of  carrier spaces  characterized by the wavefunctions
satisfying the boundary condition 
\be
\Psi(r,\theta+2\pi) = e^{i2\pi \al} \Psi(r,\theta), \qquad 0\leq \al <1.
\ee
So, we have transferred the parameter $\rho$, labeling the classes of gauge potentials, to a phase
$e^{i2\pi\al}$ of the wavefunctions. For any $\al$ we
have a differential realization of the Lie algebra $\algebra$ with an invariant gauge
potential, but  not any more a realization of the group
$\group$. Nevertheless, we obtain a solvable system, whose eigenfunctions are the
so called ``moving states'', which  have interest in the interpretation of the Hall
effect \cite{avron}.
Now, the equation for the spectrum keeps the same form as (\ref{hamiltonian2}) or
equivalently  (\ref{abc}) but  where the parameter $m$  must be substituted by $m+ \al$.

In this
case all the considerations about the ladder operator method also remain valid
with the same annihilation lines except that the physical vacuums on these
lines are parametrized by $m= n+ \al$, $n\in \Z$, maintaining the restrictions
(\ref{1})--(\ref{3}). However, there are some properties that have changed
drastically.  When $\al =0$ each physical sector is invariant under the 
action of the operators $\{J^{\pm},A^{\pm}\}$ (for $\k<0$, in a two-fold way),
and each state is connected with any other by means of such operators as we seen in the above
subsection. But when
$\al \neq 0$ each physical sector is broken into subsectors with the following
modifications: i) the states of each subsector are linked by means of
$\{J^{\pm},A^{\pm}\}$, but states belonging to different subsectors are not connected any
more. ii) Each subsector is not invariant under all the shift and ladder
operators, so that the action of some of them lead to non physical states.
These properties are illustrated separately in Fig.~3 for 
$\k>0$ and  $\k<0$.  Note that all these features can be obtained, of course, in
the frame of the hypergeometric equation of subsection \ref{eigenfunctions}, 
but there we would loose the operators that are so useful in describing the change in the
spectrum.

From Figs.~3 and 4 one can understand easily the `index' \cite{avron} associated to each energy
level. This index is defined as the difference between the number of states that joint and  leave
an energy level when the parameter $\alpha$  increase by a period. For
$\k\leq 0$ the index is 1, but in the compact case ($\k >0$)  the index is 0.
\pagebreak

\
\vskip-2.5cm
\ 
\begin{center}
\epsfig{file=Landau4.eps,width=13.cm,height=11.7cm }
\end{center}
\
\vskip-1.5cm
\ 
\begin{center}
\epsfig{file=Landau1.eps,width=13.cm,height=9.8cm }
\end{center}
\
\vskip-1.5cm
\ 
\begin{center}
\footnotesize{Fig.~1:  Physical sectors in grey for $\k>0$ (top), and $\k<0$
(botton), together with annihilation lines for $A^-$ (solid lines) and $A^+$ 
(dashed lines).}
\end{center}
\pagebreak
\

\

\begin{center}
\epsfig{file=Landau5.eps,width=16.8cm,height=8.4cm }
\end{center}
\ 
\ 
\begin{center}
\epsfig{file=Landau2.eps,width=14cm,height=9.2cm }
\end{center}

\medskip

\begin{center}
\footnotesize{Fig.~2:  Lattice of normalizable states for $\k>0$ (top), 
and $\k<0$ (botton).}
\end{center}
\pagebreak

\ 
\vskip-2cm
\ 
\begin{center}
\epsfig{file=Landau6.eps,width=17.2cm,height=8.4cm }
\end{center}
\

\

\begin{center}
\epsfig{file=Landau3.eps,width=14cm,height=9.2cm }
\end{center}

\begin{center}
\footnotesize{Fig.~3:  Lattice of normalizable states with aperiodic boundary
conditions $(\al \neq 0)$ for
$\k>0$ (top),  and $\k<0$ (botton).}
\end{center}

\sect{Remarks and conclusions\label{conclusiones}}

 In order to interpret physically our results, we will consider  dimensions
for the Hamiltonian  $H_\k = {\overline{C}}/2$ with the Casimir given in
(\ref{casex}). By means of a multiplicative factor and the identification
\be
\bb = \tau q \frac{|{\bf B }|}{\hbar c},
\ee 
where $q$ is the charge of the physical system, $|{\bf B}|$ is the intensity of the magnetic
field and $\tau=+1$ or $-1$ indicates that the magnetic field points in or out the direction of
$J_{12}$, we can rewrite
$H_\k$  in the form
\begin{equation}\label{landauehamiltoniano}
H_\k=\frac{\hbar ^2}{2m_0}(J_{01}^2+J_{02}^2+\k J_{12}^2)
- \tau \frac{\hbar q}{m_0}|{\bf B}|J_{12}.
\end{equation}
The  spectrum of the Hamiltonian (\ref{landauehamiltoniano}) is  
\be\label{energia1}
 E_{l}^{\k}=\frac{|q|\hbar}{ m_0 c}|{\bf B}|(l+\frac{1}{2})+  
\frac{\hbar ^2 \k }{2m_0 } l(l+1) .
\ee 
Note that we have put together expressions (\ref{eigenv}) and (\ref{eigenv2}).  The second term of
the energy (\ref{energia1}) has a marked geometric meaning since it depends on $\k$ wich is
related with the curvature radius, $R$, of the configuration space ($|\k| = 1/R^2$).  In the limit
$\k \to 0$ this term disappears as it is the case in the euclidean plane. 
 
When $\k \neq 0$ we can consider  a monopole with  an  associated radial 
magnetic field $\bf B$.  Dirac quantization condition \cite{dirac} gives that 
\be
 |{\bf B}|=\frac{\hbar n}{|e|R^2}=\frac{\hbar n |\k| }{|e|},
\ee
with $n$  a natural number and $e$ the  elementary negative charge. Remark that if we want that 
in the limit $\k \to 0$  the field $\bf B$ remains finite, 
$n \to \infty$  in order to keep the term $\k n$ constant.

Now the spectrum is 
\be
E_{l}^{n,\eta}=\frac{|q|\hbar}{ m_0 c}|{\bf B}|(l+\frac{1}{2})+  \eta
\frac{ \hbar |e|  |{\bf B}| }{2 m_0 c n} l(l+1) , 
\ee
with $\eta=+1$ or $ -1$ according to the configuration space has postive or negative curvature,
respectively. 

The Landau systems considered in this paper can be obtained from the 
relevant symmetry groups of the involved magnetic fields by means of 
their  local realizations. Each irreducible component inside a class of local
realizations has labels  $(\bb,l,\k)$ whose meaning is the following:
(i) a real parameter $\bb$ proportional to the intensity of an external
magnetic field  interacting with the
quantum system (this intensity is quantized for $\k\neq 0$,
which implies the quantization of the magnetic charge); (ii) a positive integer
$l$ that determines the energy of the Landau level and characterizes the
bounded (discrete) representations of the magnetic groups; and (iii) a real 
label $\k$, measuring the curvature of the two dimensional configuration
space, whose standard values $\k=1,0,-1$ correspond to the three Landau
systems, spherical, planar and hyperbolic, respectively. The expressions in
terms of $\k$ facilitates the comparison among these systems showing their
analogies and differences as well. Besides this, such expressions have sense
for any real value of $\k$. This property tell us how to connect correctly the
spherical and/or hyperbolic systems to the well known planar Landau system
by the contraction procedure $\k \to 0$. Remark, as we mentioned before,
that $\k$ also shows how the geometry of the  configuration space
contributes to the energy eigenvalues by means of the term $\k l(l+1)$.

Summarizing, we have attached a physical meaning to all the parameters
labeling each class of local realizations up to gauge equivalence. Let us
insist here on the role played by the local character of this
classification. Once fixed $\k $, there are several choices for the
remaining parameters, $\bb$ and $l$, giving rise to (globally) equivalent
irreducible representations. For instance, if $\k>0$, as long as $l {+}
\bb/\k=j$, with $j$ being a fixed positive half-integer we will always get
$(2j{+}1)$--dimensional equivalent irreducible representations of $SU(2)$, the
universal covering of $SO(3)$.  However, different elections of the pairs
$(\bb,l)$ fulfilling such a condition do not belong to the same class of {\em
local equivalence}. This is the reason why they should be considered as
describing non-equivalent physical systems, e.g., systems with different
energy ($l$) evolving under a different magnetic field ($\bb)$. In
mathematical terms we would say that, for $\k>0$, the hypergeometric
functions include in {\em several ways} each representation of 
$SU(2)$. 
To see the meaning of this point more explicitly take, for instance,
$\k=1$, and make the following two choices: (1)  $(l=j,\bb=0)$; and (2)
$(l=0,\bb=j)$.  In the first case the Landau energy level is
$E_{(1)}=(j+1)j/2$, while in the second one $E_{(2)} = j/2$, so both 
systems are disequivalent from a physical point of view. Figure~4 
displays the density probability corresponding to each of the five
eigenfunctions for these two cases when $j=2$. This is another way to
make explicit the local inequivalent of both sets of eigenfunctions.
\bigskip

\hskip-1cm
\epsfig{file=Landau10.eps,width=8cm,height=5cm }
\epsfig{file=Landau8.eps,width=8cm,height=5cm }

\begin{center}
\footnotesize{Fig.~4: Density probability $|R_{l,m}^{\bb,\k{=}1}(r)|^2$ of the
eigenfunctions $m{=}0,\pm 1,\pm2$ 
($|R_{l,m}^{0,\k{=}1}(r)|^2 {=} |R_{l,-m}^{0,\k{=}1}(r)|^2$)  \\
in the representation  $l{=}2,\bb{=}0$ (left), and those with
$m{=}0,\dots,4$ for $l{=}0,\bb{=}2$  (right); $r\in [0,\pi]$. }
\end{center}

It is interesting also to see how the initial  eigenfunctions
starting from the sphere $S_{\k=1}\equiv S^2$ evolve into those defined in
the plane $S_{\k=0}\equiv {\mit\Pi}^2$. We illustrate this behavior 
for the radial density of the wavefunctions characterized by $l=0$,
$m=0,\dots,4$ and the values $\k=1$, $r\in [0,\pi]$ in Fig.~4 (right), 
$\k=1/2$, $r\in [0,\sqrt{2}\pi]$ in Fig.~5 (left), and 
$\k=0$, $r\in [0,+\infty)$ in Fig.~5 (right).
\bigskip

\hskip-1cm
\epsfig{file=Landau9.eps,width=8cm,height=5cm }
\epsfig{file=Landau7.eps,width=8cm,height=5cm }

\begin{center}
\footnotesize{Fig.~5: Density probability $|R_{0,m}^{2,\k}(r)|^2$ of
the eigenfunctions $m{=}0\dots,4$ of the representations \\
$l{=}0,\bb{=}2$ for $\k=1/2$, $r\in [0,\sqrt{2} \pi]$ (left), and for
$\k=0$, $r\in [0,+\infty)$ (right).}
\end{center}

The continuous spectrum of the hyperbolic Landau quantum systems can be easily understood
through horocyclic coordinates that allow to transform these hyperbolic systems into Morse
systems. These coordinates display $R$--separability, however they contract to cartesian
coordinates on the plane.

From the symmetry group we have derived operators, that leave invariant each
eigenspace,  but also we have obtained ladder
operators $\{A^{\pm}\}$ (not related to the space symmetry) linking
consecutive  eigenspaces $\varepsilon_\k^l \to  \varepsilon_\k^{l\pm 1}$.
These new operators act in the same lattice of physical states defined by the
group generators $\{J^{\pm}\}$. The main interest of the ladder operators is
that they show the link of Landau systems with oscillators on curved
speces and  allow to understand in a simple way the spectrum for
quasi-periodic wavefunctions or moving states.

\vskip 1cm


\noindent{\Large \bf Appendix}
\appen

\subsection*{A.  Local realizations of symmetry groups}

In QM the elements $g$ of the symmetry group $G$ of a quantum system are 
represented by local unitary operators $U(g)$ acting on the space of 
wavefunctions $\psi$ defined on the space-time manifold (a homogeneous
space of $G$) $X$ in the form 
\be\label{locreal}
 \psi'(x')\equiv (U(g) \psi)(g\,x) 
= A(g,x) \psi(x), \qquad g\in G,\ x\in X,
\ee
where $A(g,x)$ is a matrix-valued function. In the particular case of 
one-component wavefunctions $A(g,x)$ is simply a  phase function, i.e,
$A(g,x)=e^{i\zeta(g,x)}$ (in the following we will only  consider
one-component wavefunctions).  

The operators $U(g)$ (\ref{locreal}) close, in general, not a true 
representation  but a projective (or `up to a factor') representation 
of $G$ \cite{bargmann54} that, henceforth,  we shall call {\sl local
realization}, 
\be
U(g_2)U(g_1) =\omega(g_2,g_1) U(g_2g_1),\quad g_2,g_1 \in G.
\label{UU}
\ee
The function $\omega :G\times G \to U(1)$ is the factor system of the 
realization and it is a 2--cocycle, i.e., $\omega \in {\bf Z} ^2(G,U(1))$. 
It is often used the notation $\omega(g_2,g_1) =\exp\{i\,\xi(g_2,g_1)\}\in
U(1)$, where  $\xi(g_2,g_1)\in \R$ is called the exponent of $\omega$.
Only if $\omega(g_2,g_1)=1,\forall g_2,g_1 \in G$, the realization $U$
is, in fact, a  true representation.

The equivalence of local realizations must keep the local character and it 
is called  {\em gauge, or local,  equivalence}.  Given two local realizations
$U$ and 
$U'$ of $G$, they are said to be gauge equivalent if there are a function
$\lambda : G \to U(1)$  and a linear operator $T$ acting locally in the 
carrier space, i.e., $[Tf](x)=T(x)f(x)$, with $T(x)$ a phase factor, such
that 
\be\label{le0}
U'(g)= \lambda(g) T U(g) T^{-1}, \qquad  \forall g \in G.
\end{equation}
Their corresponding phase functions are related by
\begin{equation}\label{le}
e^{i\zeta'(g,x)}=\lambda(g) T(gx) e^{i\zeta(g,x)} T^{-1}(x).
\ee
The factor systems $\omega$ and $\omega'$ associated to two equivalent 
realizations, $U$ and  $U'$ of $G$,  are said to be equivalent; they satisfy
\be
\omega '(g_2,g_1)=
\lambda^{-1}(g_2,g_1)\lambda(g_2)\lambda(g_1)\omega(g_2,g_1), 
\quad \forall g_2,g_1 \in G .
\ee
In particular, a factor system is trivial if it is equivalent to
$1$; in other words, it is a 2--coboundary,  
$\omega \in {\bf B} ^2(G,U(1))$.
The quotient
${\bf H} ^2(G,U(1))={\bf Z} ^2(G,U(1))/ {\bf B} ^2(G,U(1))$ 
is the second cohomology group of $G$ and it takes part on the
characterization of the equivalence classes of the unitary irreducible
projective representations of $G$. 

The  classification  of all the local realizations up to gauge equivalence
has been solved in general terms in \cite{mado83}. 
As a first step the local realizations are linearized, e.g., instead of
computing directly  the representations up to a factor of $G$ we can
get them from the linear local  representations of a new group 
$\overline G$, which is a central extension of  $G$ by an Abelian group
$A$. It can be shown that $A$ is the dual of (a subgroup of) the second
cohomology group of $G$, $\hat{{\bf H}}^2(G,U(1))$. Therefore, a necessary
ingredient is  ${\bf H}^2(G,U(1))$ whose computation can be done by
solving the equivalent problem of the central extensions of $G$ by
$U(1)$ \cite{azcarraga}. 
The local representations $\overline{ U}$ of 
$\overline G$ originate the local realizations $U$ of $G$ once a section 
$s: G\to \overline G $, has been chosen \cite{cos84}
and provided that $\overline{U}|_{\hat{{\bf H}}^2(G,U(1))}\subset U(1)$.
Then, the realization $U$ associated to the representation
$\overline{ U}$ is given by  
\be
{U}(g) :=\overline{U}(s({g})).
\ee
If $G$ is simple (such as it is $SO(3)$) the representation group $\overline G$
is simply the universal covering group ($SU(2)$ in this case).

\subsection*{ B.  Central extensions of Lie groups}

A group $\overline G$ is a central extension of $G$ by the Abelian
group $A$ if in the following exact sequence of group homomorphisms
$$
1 \rightarrow A\stackrel{i}
\rightarrow {\overline G}\stackrel{p}
\rightarrow G\rightarrow1\, ,
$$
 $A$ is in the center of ${\overline G}$ and $G = {\overline G}/A$. The
search of central extensions is a cohomologic problem, since each
extension of $G$ by $A$ has associated an element $w\in {\bf Z}^2(G,U(1))$.
It can be shown that there is a bijection between the equivalence classes
of central extensions of $G$ by $A$ and ${\bf H}^2 (G,A)$. 

We can study the central extensions of Lie groups (at least, in a
neighborhood of the identity element; global properties must be considered
separately taking into account the universal covering group) in terms of
central  extensions of their Lie algebras.  As in  this work  we are interested
in extensions,
${\overline G}$, of a Lie group $G$ by $U(1)$, so the Lie algebra
$\overline{{\cal G}}$ (of 
${\overline G}$) is an extension of ${{\cal G}}$ (the Lie algebra of $G$) by
$\R$ (Lie algebra of $U(1)$). The classes of such extensions are given by
the second cohomology group ${\bf H}^2({\cal G},\R)$. We shall see now how
to build these classes.

Let $\{ X_i\}_{i=1}^ n$ be the generators of  ${\cal G}$ with commutators
\be\label{Lieequis}
[X_i,X_j]=c_{ij}^k\,X_k .
\ee
The Lie algebra $\overline{\cal G}$ will have dimension $n{+}1$, with 
generators  $\overline{X}_i$  (corresponding to each $X_i$ of ${\cal G}$)
plus a new central generator  $I$, i.e.,  $I$ commutes with all the
$\overline{X}_i$'s. The Lie commutators of $\overline{\cal G}$ will be
\be\label{nuevasLieequis}
[\overline{X}_i,\overline{X}_j]=c_{ij}^k\,\overline{X}_k+\la_{ij}I ,
\ee
where the  structure constants   $c_{ij}^k$  are the same of 
(\ref{Lieequis}) and $\la_{ij}\in \R$. Since relations 
(\ref{nuevasLieequis}) have to define a Lie algebra  the  parameters
$\la_{ij}$   are not arbitrary; they must verify the conditions obtained by
imposing the Jacobi identity.  

Two extensions given by the sets $\{\lambda_{ij}\}$ and
$\{\lambda'_{ij}\}$, are said to be equivalent if they are related by a
change of basis
${\overline X}_i\to {\overline Y}_i={\overline X}_i+\mu_i I$,  with
$\mu_i\in \R$. Then, the commutators (\ref{nuevasLieequis}) become
\be\label{Lieies}
[{\overline Y}_i,{\overline Y}_j]
= c_{ij}^k\,{\overline Y}_k+(\la_{ij} -c_{ij}^k\,\mu_k)I . 
\ee
Therefore, the equivalence of the two sets of extension parameters take
the form
\be
\lambda'_{ij} = \la_{ij} -c_{ij}^k\,\mu_k  \, .
\ee
If we can find a set $\{\mu_k\}$, such that $\lambda'_{ij} =0$, we will say
that the extension determined by  $\{\lambda_{ij}\}$ is trivial. In such a
case $\overline{\cal G}={\cal G}\oplus \R$, and the corresponding
extension group $\overline{G} = G\otimes U(1)$ is a direct product.

In our particular case, $\group$ has only a nontrivial extension when
$\k=0$  described by the commutator $[J_{01}, J_{02}]=\lambda I$,
the other two commutators are nonzero and the possible
extension parameters $\la_{ij}$ related with them can be reabsorbed by an
equivalence. In conclusion, ${\bf H}^2(so_{\k=0}(3),\R)=\R$, and
${\bf H}^2(so_{\k}(3),\R)=0$ if $\k\neq 0$. In any case, even when  $\k\neq 0$,
we shall take into account the (trivial) extension
$[J_{01}, J_{02}]=i\k J_{12}+\lambda I$, in order to have a common
formalism for any $\k$-value. The group $\sgroup$ used to build the
realizations of $\group$ according to Appendix A will be referred to as the
`magnetic group'. Its  Lie algebra, $\salgebra$, is an extension of $\algebra$ by $\R$, 
whit Lie commutators  (now the central element is called $B$)
\be \label{extended}
[\overline{J}_{01},\overline{J}_{02}]=i\kappa\, \overline{J}_{12} + i{B},
\quad  [\overline{J}_{12},\overline{J}_{01}]=i\overline{J}_{02},\quad 
[\overline{J}_{12},\overline{J}_{02}]=-\overline{J}_{01} , \quad  [.,B]=0 \, . 
\ee

\subsection*{ C.  Local representations of the magnetic groups}

Now we are in conditions to characterize the classes of local realizations 
of  ${ SO}_\k(3)$ up to gauge equivalence in terms of the local
representations of $\overline{ SO}_\k(3)$.

\noindent
{\bf Theorem}. {\it The  local 
realizations,  $U_{\la,\bb}$, of $SO_\k (3)$ are obtained by means of the
representations  of $\overline{SO}_\k(3)$ induced from the one-dimensional representations 
\begin{equation}\label{induction}
D_{\la,b}(\phi,\zeta) = e^{-i(\la \phi + b \zeta )},
\qquad  b ,\la \in {\R}
\end{equation}
of the abelian isotropy subgroup of 
$x_0$, generated by $\overline{J}_{12}$ and $B$. }
\medskip

The induced representations  can be straightforwardly computed 
(see Ref.\   \cite{COS85} where the local realizations for the Euclidean
group are worked out). So, we  supply  below the representations of the
$\salgebra$ generators (\ref{locgen}), with $\k \neq 0$, obtained by
induction from (\ref{induction}). 
\bea\label{fields} 
&&J_{01}=J_{01}(r,\theta) - (\la - \frac{b}{\k} \cos\sqrt{\k} r)
\frac{\sqrt{\k}\sin\theta}{\sin\sqrt{\k}\, r},\nonumber\\  
&&J_{02}=J_{02}(r,\theta) +(\la - \frac{b}{\k} 
\cos\sqrt{\k} r)\frac{\sqrt{\k}\cos\theta}{\sin\sqrt{\k}\, r},\\  
&&J_{12}=J_{12}(r,\theta), \qquad B=-b.\nonumber
\eea
Here, some remarks concerning to expressions (\ref{fields}) are
appropriate:

\begin{itemize}

\item[(i)] $\k \neq 0$. The label $\la$ in the representation (\ref{induction})
of $\overline{SO}(2)$ must be a half integer because we work with the
universal covering of ${SO}_{\k}(3)$, which has center $\Z_2$. The value of
$\la$ determines a class of local equivalence, while $b$ is an irrelevant
real parameter that can be gauged away. However we can choose $b$ in an
appropriate way: if we take $\la = {b}/{\k}\equiv \bb/\k$, the generators
(\ref{fields} ) become differentiable in the north pole. A second reason for
this choice is given below.

\item[(ii)] $\k = 0$. In this case  $\la$ is irrelevant (it can be
arbitrarily changed by means of a pseudoequivalence (\ref{le0})), but the
parameter $b\in \R$ now becomes significative and determines the class of
local realization.  If we look at the realization (\ref{fields}) of the
generators for $\k \neq 0$ we see that it has not a well defined limit when
$\k \to 0$. If we want to get in this way the correct nontrivial expressions
for $\k=0$ we must choose $\la = {b}/{\k}\equiv \bb/\k$, as said above. With this
choice we have only one parameter, $\bb$, that determines the local
class for any $\k$. This is the final result shown in (\ref{campos-Oexb}).
\end{itemize}

\subsection*{ D.  Fiber bundles and local representations}

Let us consider the principal bundle 
$\sgroup (\sphere\, ,\pi,SO(2)\otimes \R)$,
with total space  $\sgroup$, base space  $\sphere$, projection 
$\pi:\sgroup \to \sphere$, and structure group $SO(2)\otimes \R$,
generated by $\{J_{12},B\}$. Each irreducible one-dimensional
representation, $D_{\la,\bb}$, of the isotropy subgroup, $SO(2)\otimes
\R$, of $x_0$  allows us to build up an associated vector bundle,
$E_{D_{\bb,\la}}(\sphere\, ,\pi_{E},\C)$, whose 
fiber is the support space, $\C$, of $D_{\bb,\la}$, 
where $\sgroup$ acts in a natural way. This action on
the  vector bundle translated to the linear space of bundle  sections, i.e.,
Borel maps $f: \sphere\to E$ (which may be  identified with the
wavefunctions of Section 3) defines the induced local representations, and
the  restriction to $\group$, by means of the section $s : \group \to \sgroup$,
gives the local realizations. The gauge equivalence defined in the wavefunction space
has its counterpart in terms of automorphisms of the principal bundle
\cite{asorey83,no92}.

It can be shown \cite{no92,hno94} that there is an invariant connection, 
$\Theta$, under the action of $\sgroup$  on the principal bundle 
$\sgroup (\sphere\, ,\pi,SO(2)\otimes \R)$. 
The pull-back of $\Theta$ on the base space $\sphere$ is
represented on the associated vector bundle by the one-form 
$A=A_{\mu}(x)dx^{\mu}$. The invariant condition expressed in terms of $A$
is \cite{hno94},
\be\label{pullbackequation}
{\cal L}_{X} A -i d\, W=0, \qquad \forall X \in {so}_\k(3) ,
\ee
where ${\cal L}_{X}$ denotes the Lie derivative of  the vector field $X(x)$.
Making use of the fields (\ref{fields}) we arrive at the potential
\begin{equation}\label{vector-potential} 
A_{r}=0,\qquad  A_{\theta}= b/\k -\la \cos\sqrt{\k} r .
\end{equation}
The limit $\k \to 0$ is not defined but, if in agreement with the
considerations of Appendix C, we take $\la = {b}/{\k}\equiv \bb/\k$ we obtain
a well behaved potential (\ref{potential}). 

A significant property of an invariant connection on a principal bundle is 
the following one.  Let us write the Lie commutators of the
abovementioned basis (\ref{fields0}) of the Lie algebra $\algebra$
in the form
\begin{equation}\label{ec}
[X_{i},X_{j}]=c^{k}_{ij} X_{k},
\end{equation}
where the  structure constants $c^{k}_{ij}$ are given in (\ref{kso3}).  
Let $\overline{X}_{i}$  be the implementation of the vector fields 
${X}_{i}$ of ${so}_\k(3)$  as elements of  $\overline{so}_{\k}(3)$ according
to the vector field realization (\ref{campos-Oexb}). Then, if we define the
new set of generators
$X^{*}_{i}=X_{i}(x)^{\mu}D_{\mu}$ with $D_{\mu}$ 
the covariant derivatives, i.e., the horizontal  lifts of the fields
$ X_{i}=X_{i}^{\mu}(x)\partial_{\mu}$ of $\algebra$, the following
commutators are satisfied:
\begin{equation}\label{mixing}
[\overline{ X}_{i},X^{*}_{j}]=c^{k}_{ij}X^{*}_{k}, 
\ee
where the coefficients $c^{k}_{ij}$ coincide with the structure constants 
of (\ref{ec}).

The commutation relations ({\ref{mixing}}) suggest that  if
${C}_\k(X_{j})$ denotes the Casimir (\ref{casimir}) of ${SO}_\k(3)$, then 
${C}_\k(X^{*}_{i})$ is a quadratic Casimir of $\sgroup$. So, it  may differ 
with  $\overline{C}_\k({\overline X}_{j})$ in a constant, but incidentally,
in our case both coincide.

\section*{Acknowledgments}

 M.A.O. thanks to  Centro Internacional de Ciencias A.C. 
de Cuernavaca (M\'exico) for his hospitality where a part of this 
work has been concluded.  This work has been partially supported by 
DGES of the  Ministerio de Educaci\'on y Cultura of Spain under 
Project PB98-0360 and the Junta de Castilla y Le\'on (Spain).



\begin{thebibliography}{99} 

\bibitem{landau} L. Landau and E. Lifchitz, ``Quantum Mechanics:
Non-Relativistic Theory'',  Pergamon, New York, 1977.

\bibitem{dunne} G. V. Dunne, {\sl Ann. Phys.}  (N.Y.) {\bf 215} (1992)  233 .
 
 \bibitem{klitzing}  K.V. Klitzing. G. Dorda and M. Popper,   
{\sl Phys. Rev.  Lett.} {\bf 45}  (1980) 494.

\bibitem{prange90} R.F. Prange and S.M. Girvin, ``The Quantum Hall Effect'',
Springer, New York, 1990.

\bibitem{BW48} V Bargmann  and  E. P. Wigner, {\sl Proc. Natl. Acad. Sci.
U.S.}  {\bf 34}  (1948) 211.

\bibitem{H76} H. Hoogland, 
{\sl Il Nuovo Cimento} {\bf B 32}  (1976) 427. 

\bibitem{COS85} J.F. Cari\~nena, M.A. del Olmo and M. Santander, 
{\sl J. Math. Phys.} {\bf 26}  (1985)  2096. 

\bibitem{olmo01}   J. Negro, M.A. del Olmo and A. Rodr\'{\i}guez-Marco,  {\em Group theory and
Landau quantum systems} in Publ. de la RSME vol. II, p. 223,  (M.A. del Olmo and  M. Santander ed.),
RSME, Madrid, 2001.

\bibitem{NO90} J. Negro and M.A. del Olmo,
{\sl J. Math. Phys.} {\bf 31}  (1990)  2811. 

\bibitem{wigner39}   E. P. Wigner,  {\sl Ann. Math. } {\bf 40}  (1939) 149.

\bibitem{ball93} A. Ballesteros, F.J. Herranz, M.A. del Olmo and M. Santander, 
{\sl J. Phys. A: Math. Gen.} {\bf 26}  (1993) 5801. 

\bibitem{olmo2}  M. Santander, F.J. Herranz and M.A. del Olmo, 
{\sl Kinematics and homogeneous spaces for symmetrical contractions of
orthogonal groups} in {``Anales de F\'{\i}sica, Monograf\'{\i}as''}
1, vol. I.  p. 455, (M.A. del Olmo, M. Santander and J.
Mateos--Guilarte ed.),  CIEMAT/RSEF,  Madrid, 1992.

\bibitem{inonu} E. In\"on\"u and  E. P. Wigner,  {\sl Proc. Natl. Acad. Sci.
U.S.} {\bf 39}  (1953) 510; E. In\"on\"u, {\sl  Contractions of Lie
groups and their representations}, in  {``Group theoretical concepts
in elementary particle physics''}, (F. G\"ursey ed), Gordon and Breach, 
New York, 1964.

\bibitem{aops97} J.A. de Azc\'arraga,	 M.A. del Olmo,  J.C. P\'erez-Bueno and M.
Santander,  {\sl J. Phys. A: Math. Gen.}  {\bf 30}  (1997) 3069.

\bibitem{herranz} F.J. Herranz, {``Classical and quantum  Cayley-Klein groups''}. 
Ph. D. Thesis, Universidad de Valladolid, 1995.

\bibitem{bargmann54} V.  Bargmann,  {\sl Ann. Phys.} (N.Y.) {\bf 59}  (1954) 1.

\bibitem{Abr72}
M. Abramowitz and I.A. Stegun, {``Handbook of Mathematical Functions''},
Dover, New York, 1972.

\bibitem{dirac}   P.A.M. Dirac,  {\sl Proc. R. Soc. London, 
Ser. A} {\bf 133}  (1931) 60.

\bibitem{arratia97} O. Arratia and  M.A. del Olmo,  {\sl Fortschr. Phys.} {\bf 45}  (1997) 103.

\bibitem{olver} F.W.J. Olver, {``Asymptotics and special functions''}, 
Academic Press, New York, 1974. 

\bibitem{miller} W Miller, {``Lie Theory and Special Functions''}, 
Academic Press, New York, 1968.

\bibitem{olevsky}  M.P. Olevsky, {\sl Mat. Sborn.} {\bf 27} (1950) 379.

\bibitem{duru} I.H. Duru, {\sl  Phys. Rev. D} {\bf 28} (1983) 2639.

\bibitem{comtet} A. Comtet, {\sl  Ann. Phys. } {\bf 173} (1987) 185.

\bibitem{almorox}   A. L\'opez Almorox and C. Tejero Prieto,  {\em Symplectic reduction on the
hyperbolic Kustaanheimo--Stiefel fibration} in Publ. de la RSME vol. II, p. 153,  (M.A. del
Olmo and  M. Santander ed.), RSME, Madrid, 2001.

\bibitem{tejero}    C. Tejero Prieto ,  {\em Spectral geometry 
of the Landau problem and automorphic forms} in Publ. de la RSME vol. II, p. 287, (M.A. del
Olmo and  M. Santander ed.), RSME Madrid, 2001.

\bibitem{boyer} C.P. Boyer, E.G. Kalnins and P. Winternitz,  {\sl J. Math. Phys. }
{\bf 24} (1983) 2022.

\bibitem{higgs} P.W. Higgs, {\sl J. Phys. A} {\bf 12} (1979) 309. 

\bibitem{pogosian} Ye M. Hakobyan, G.S. Pogosyan, A.N. Sissakian and S.I. Vinitsky,
{\sl Phys. Atom. Nucl.} {\bf 62} (1999) 623.

\bibitem{natalie} J. Beckers, Y. Brihaye and N. Debergh, {\sl J. Phys. A} {\bf 32}
(1999) 2791. 

\bibitem{olmo96}  D.J. Fern\'andez C.,  J. Negro and M.A. del Olmo,
{\sl Ann. Phys.}, {\bf 252} (1996) 386.

\bibitem{refined} J. Negro, L.M. Nieto and O. Rosas-Ortiz,
{\sl J. Phy. A}, {\bf 33} (2000) 7207.

\bibitem{asorey} M. Asorey, 
{\sl J. Geom. Phys.} {\bf 11}  (1993) 63. 

\bibitem{avron} J.E. Avron and A. Pnueli, ``Landau Hamiltonians on Symmetric
Spaces" in {``Ideas and methods in mathematical analysis, stochastics, and
applications''}, vol. II (S. Alverio {\sl et al} ed.), Cambridge Univ. Press, Cambridge, 1992;
 J.E. Avron, R. Seiler and B. Simon, {\sl Phys. Rev. Lett.} {\bf 65} (1990) 2185.

\bibitem{mado83} M. A. del Olmo,  {`` Local realizations of  Lie groups of transformations''}.  Ph. D.
Thesis, Universidad de Valladolid, 1983.

\bibitem{azcarraga} J.A. de Azc\'arraga and J.M. Izquierdo, 
{``Lie Groups, Lie Algebras, Cohomology and Some Applications in Physics''}, 
Cambridge Univ. Press, Cambridge, 1995.

\bibitem{cos84} J.F. Cari\~nena, M.A. del Olmo and M. Santander, 
{\sl J. Phys. A: Math. Gen.} {\bf 17} (1984) 309. 

\bibitem{asorey83} M. Asorey,  J.F. Cari\~nena and M.A. del Olmo, 
{\sl J. Phys. A Math. Gen.}  {\bf 16}  (1983) 1603. 

\bibitem{no92} J. Negro and M.A. del Olmo, 
{\sl J. Math. Phys.} {\bf 33}  (1992) 511. 

\bibitem{hno94} V. Hussin, J. Negro and M.A. del Olmo, 
 {\sl Ann. Phys.} (N.Y.) {\bf 231}  (1994) 211. 

\end{thebibliography}
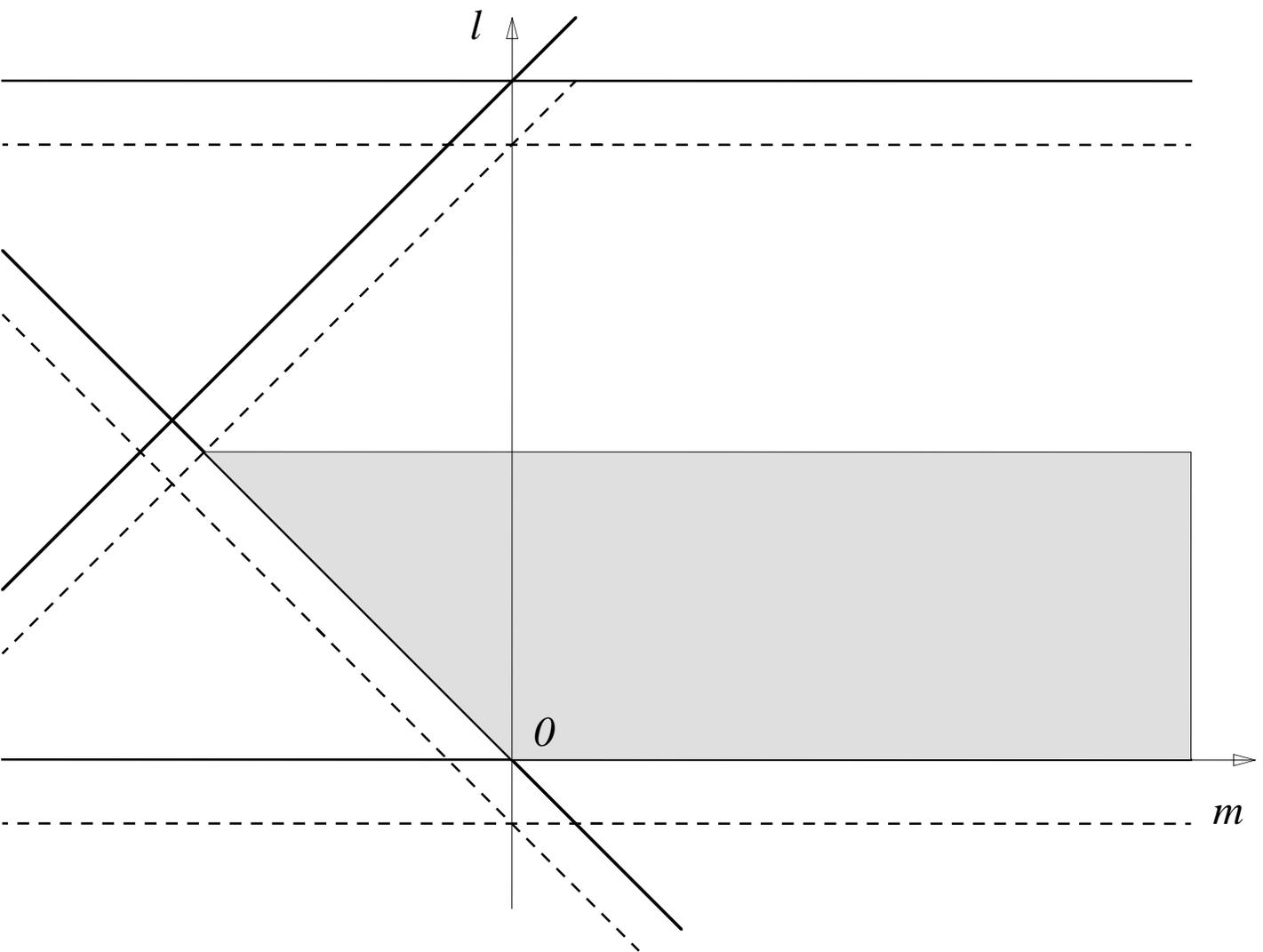
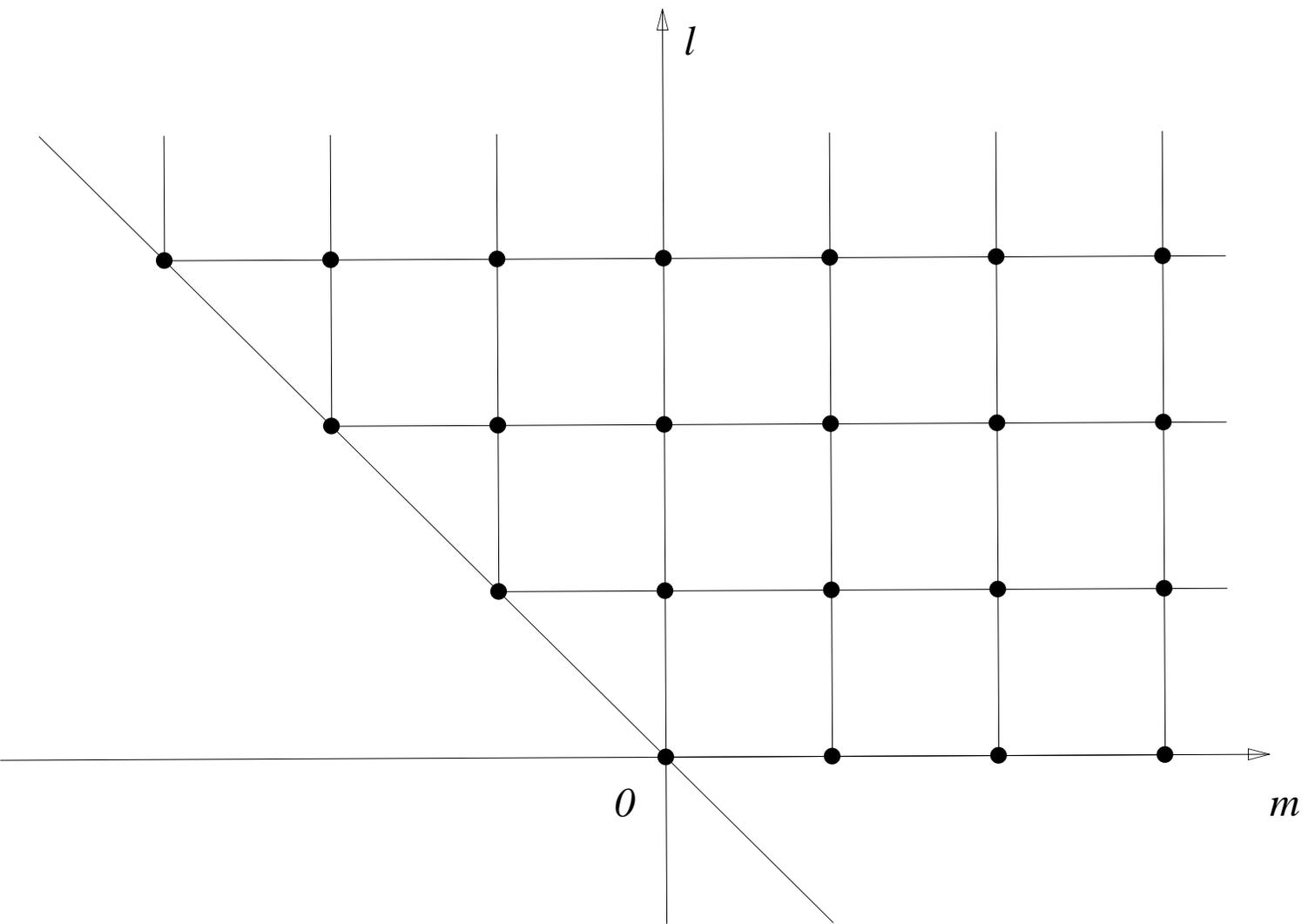
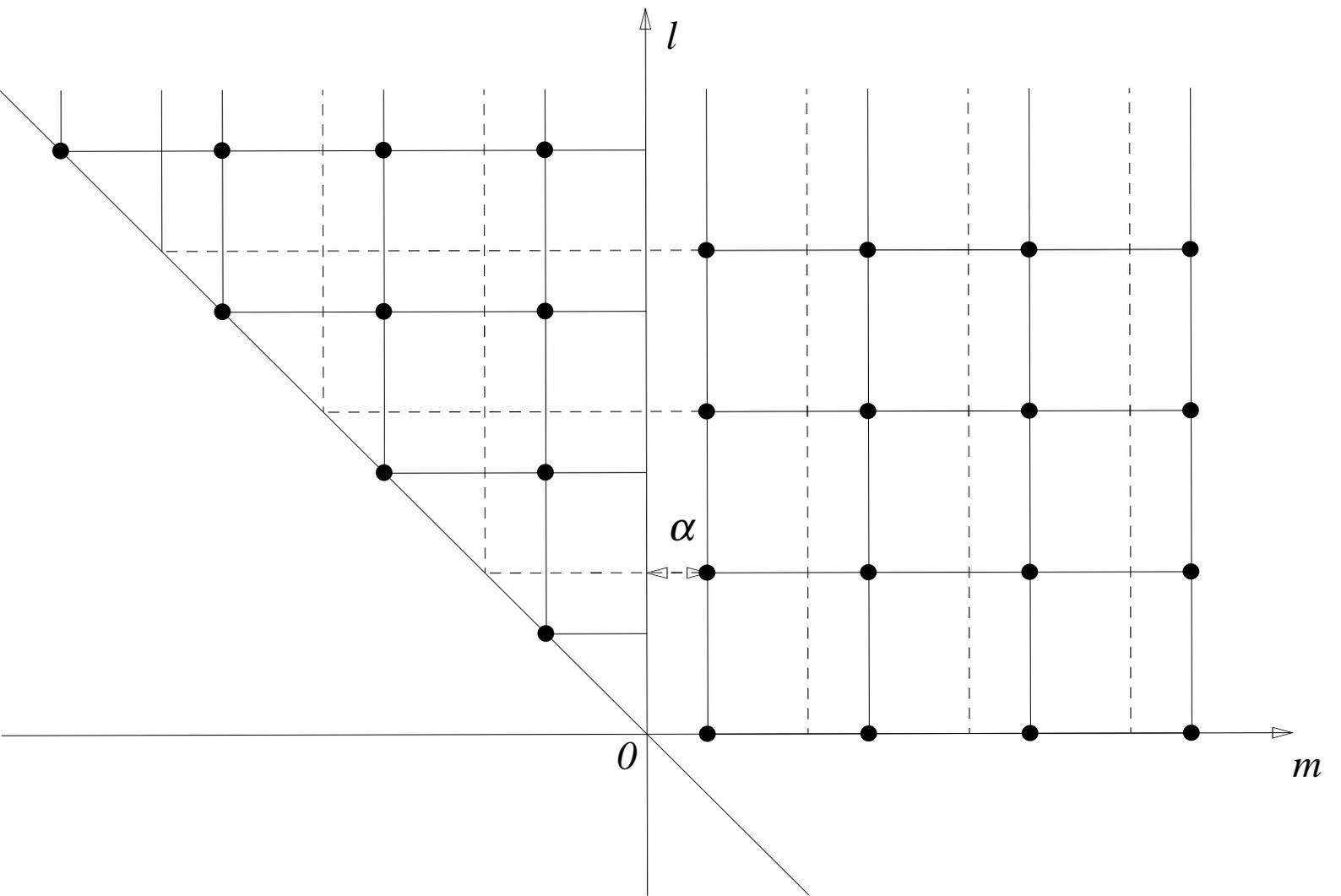
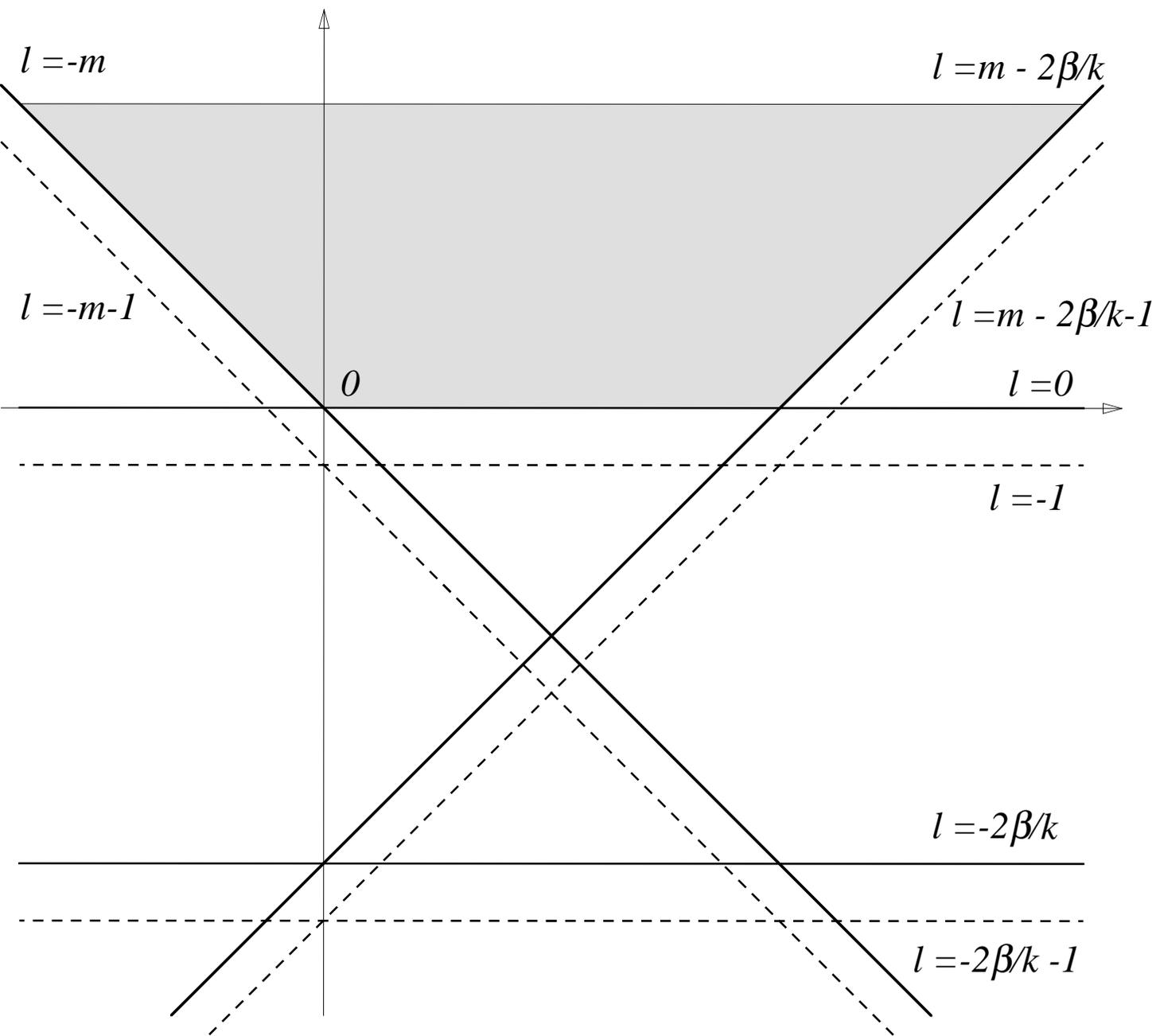
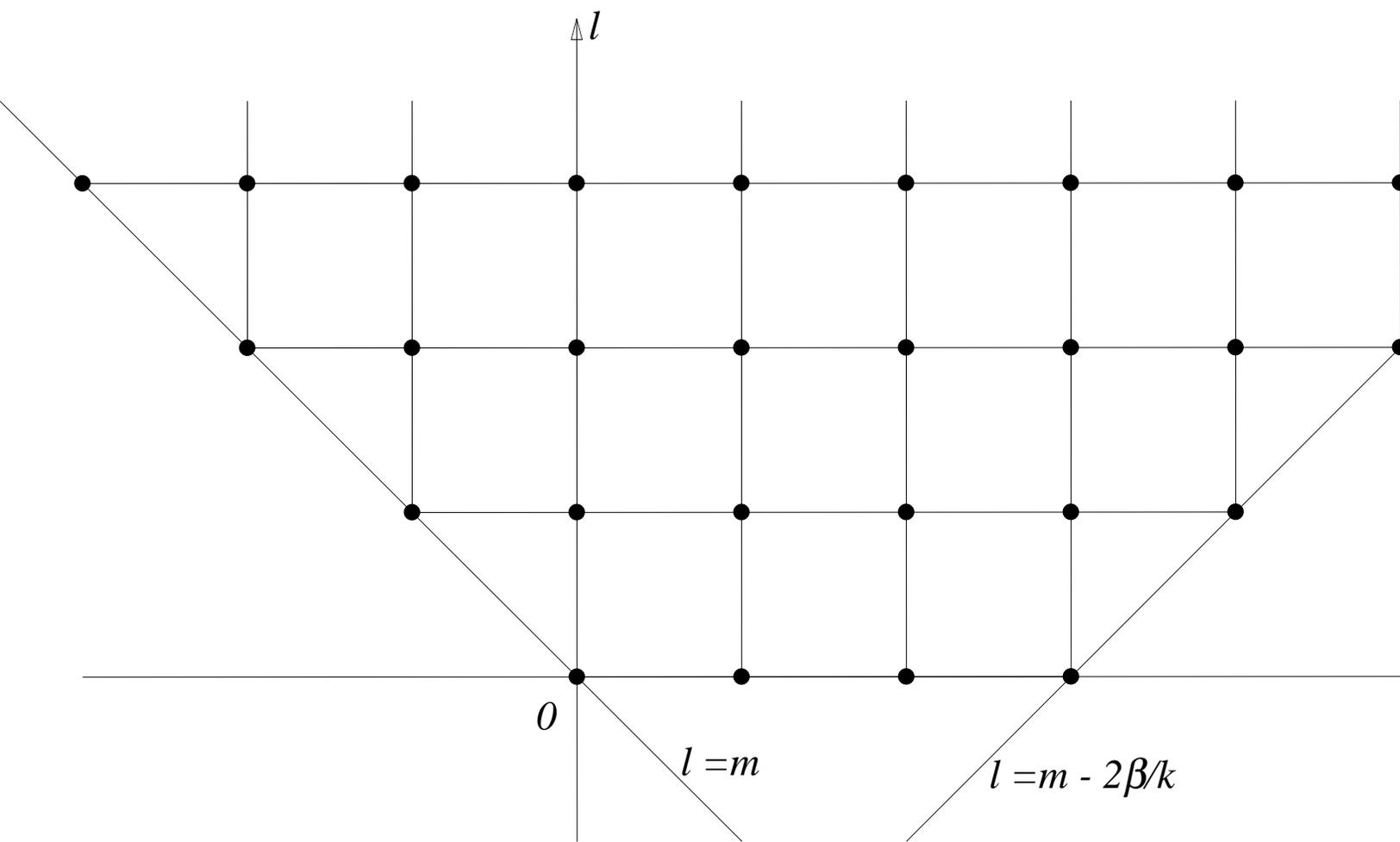
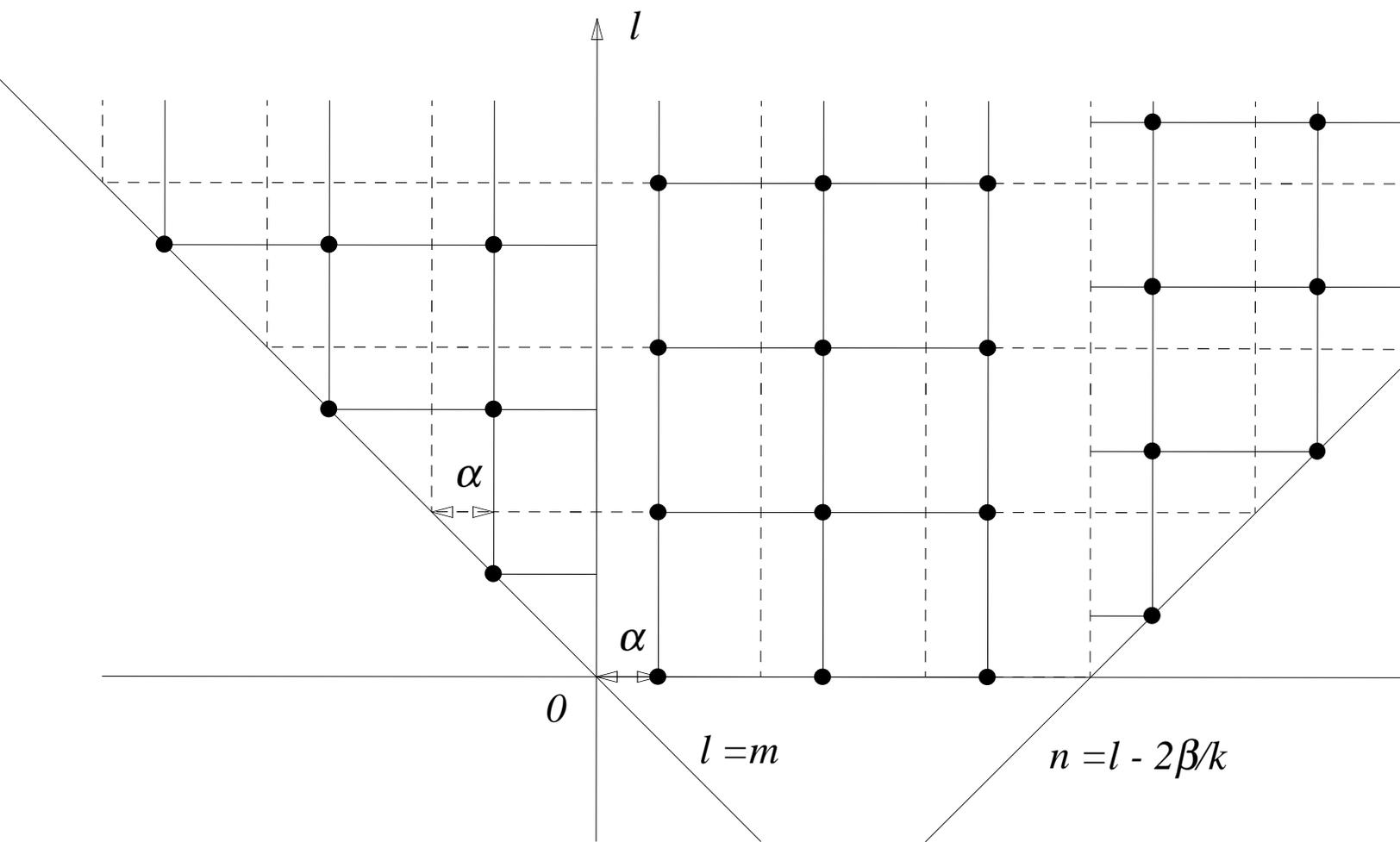
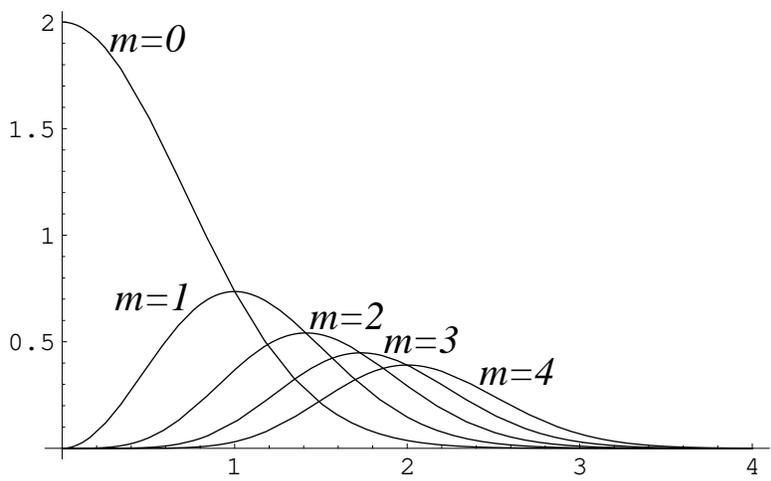
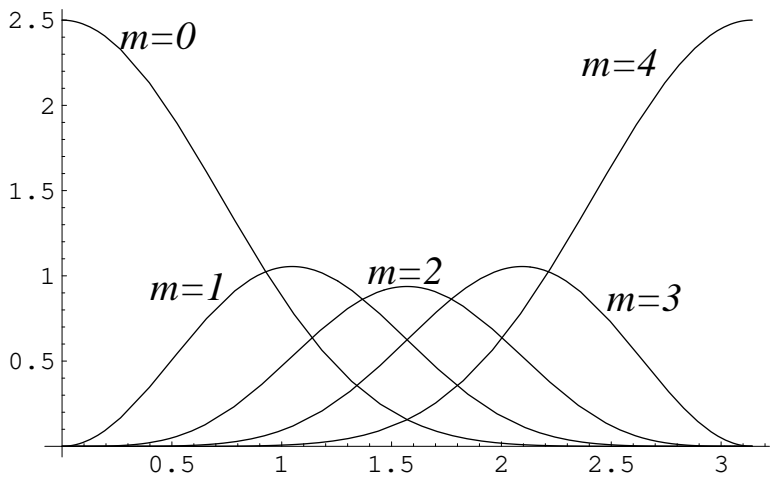
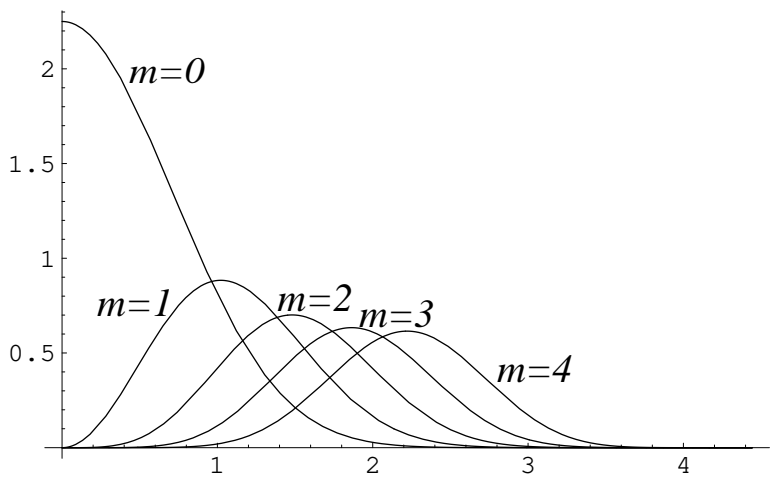
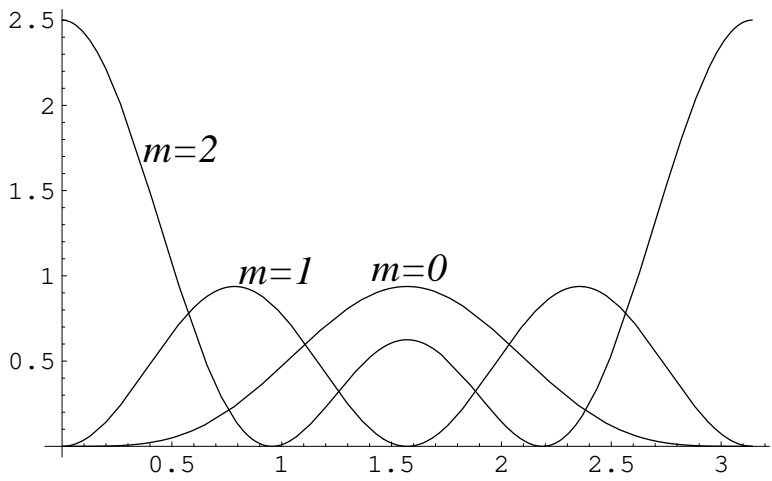
\end{document}